\documentclass[aps,prl,twocolumn,superscriptaddress]{revtex4}

\usepackage[all]{xy}
\usepackage{graphicx}
\usepackage{subfigure}
\usepackage{dcolumn} 
\usepackage{bm}      
\usepackage[active]{srcltx} 
\usepackage{color}
\usepackage{multirow}

\definecolor{dred}{rgb}{.8,0.2,.2}
\definecolor{ddred}{rgb}{.8,0.5,.5}
\definecolor{dblue}{rgb}{.2,0.2,.8}
\definecolor{dgreen}{rgb}{.2,0.5,.2}

\usepackage[colorlinks,linkcolor=dred,urlcolor=dblue,citecolor=dgreen]{hyperref}
\usepackage{amsmath}
\usepackage{amssymb}
\usepackage{stmaryrd}
\usepackage{amsfonts}
\usepackage{mathrsfs}
\usepackage{txfonts}
\graphicspath{{figs/}}

\newcommand{\Ref}[1]{(\ref{#1})}

\newcommand{\ket}[1]{| #1 \rangle}

\newcommand{\Z}{\mathbb{Z}}

\newcommand{\C}{\mathbb{C}}

\newcommand{\inv}[1]{\mathrm{Inv}_{\rm SU(2)}[#1]}

\def\be{\begin{eqnarray}}
\def\ee{\end{eqnarray}}

\newcommand{\ch}{\mathcal H}

\newcommand{\G}{\Gamma}

\newcommand{\eps}{\varepsilon}

\newcommand{\lt}{\left}
\newcommand{\rt}{\right}

\newcommand{\Ar}{\mathbf{Ar}}

\usepackage{times}

\begin{document}

\sloppy

\title{\bf Quantum Spacetime on a Quantum Simulator}

\author{Keren Li}
\thanks{These authors contributed equally to this work.}
\affiliation{State Key Laboratory of Low-Dimensional Quantum Physics and Department of Physics, Tsinghua University, Beijing 100084, China}
\affiliation{Institute for Quantum Computing and Department of Physics and Astronomy,
University of Waterloo, Waterloo N2L 3G1, Ontario, Canada}

\author{Youning Li}
\thanks{These authors contributed equally to this work.}
\affiliation{State Key Laboratory of Low-Dimensional Quantum Physics and Department of Physics, Tsinghua University, Beijing 100084, China}
\affiliation{Institute for Quantum Computing and Department of Physics and Astronomy,
University of Waterloo, Waterloo N2L 3G1, Ontario, Canada}

\author{Muxin Han}
\thanks{These authors contributed equally to this work.}
\affiliation{Department of Physics, Florida Atlantic University, 777 Glades Road, Boca Raton, FL 33431, USA}
\affiliation{Institut f\"ur Quantengravitation, Universit\"at Erlangen-N\"urnberg, Staudtstr. 7/B2, 91058 Erlangen, Germany}

\author{Sirui Lu}
\affiliation{State Key Laboratory of Low-Dimensional Quantum Physics and Department of Physics, Tsinghua University, Beijing 100084, China}

\author{Jie Zhou}
\affiliation{Perimeter Institute for Theoretical Physics, Waterloo N2L 2Y5, Ontario,
Canada}

\author{Dong Ruan}
\affiliation{State Key Laboratory of Low-Dimensional Quantum Physics and Department of Physics, Tsinghua University, Beijing 100084, China}

\author{Guilu Long}
\affiliation{State Key Laboratory of Low-Dimensional Quantum Physics and Department of Physics, Tsinghua University, Beijing 100084, China}

\author{Yidun Wan}
\email{ydwan@fudan.edu.cn}
\affiliation{Department of Physics and Center for Field Theory and Particle Physics, Fudan University, Shanghai 200433, China}
\affiliation{Collaborative Innovation Center of Advanced Microstructures, Nanjing, 210093, China}
\affiliation{Department of Physics and Institute for Quantum Science and Engineering, Southern University of Science and Technology, Shenzhen 518055, China}

\author{Dawei Lu}
\email{ludw@sustc.edu.cn}
\affiliation{Department of Physics and Institute for Quantum Science and Engineering, Southern University of Science and Technology, Shenzhen 518055, China}
\affiliation{Institute for Quantum Computing and Department of Physics and Astronomy,
University of Waterloo, Waterloo N2L 3G1, Ontario, Canada}

\author{Bei Zeng}
\email{zengb@uoguelph.ca}
\affiliation{Department of Mathematics and Statistics, University of Guelph, Guelph N1G 2W1, Ontario, Canada}
\affiliation{Institute for Quantum Computing and Department of Physics and Astronomy, University of Waterloo, Waterloo N2L 3G1, Ontario, Canada}
\affiliation{Canadian Institute for Advanced Research, Toronto M5G 1Z8,
  Ontario, Canada}
\affiliation{Department of Physics and Institute for Quantum Science and Engineering, Southern University of Science and Technology, Shenzhen 518055, China}

\author{Raymond Laflamme}
\affiliation{Institute for Quantum Computing and Department of Physics and Astronomy,
University of Waterloo, Waterloo N2L 3G1, Ontario, Canada}
\affiliation{Perimeter Institute for Theoretical Physics, Waterloo N2L 2Y5, Ontario,
Canada}
\affiliation{Canadian Institute for Advanced Research, Toronto M5G 1Z8,
  Ontario, Canada}%


\begin{abstract}

We experimentally simulate the spin networks---a fundamental description of quantum spacetime at the Planck level. We achieve this by simulating quantum tetrahedra and their interactions. The tensor product of these quantum tetrahedra comprises spin networks. In this initial attempt to study quantum spacetime by quantum information processing, on a four-qubit nuclear magnetic resonance quantum simulator, we simulate the basic module---comprising five quantum tetrahedra---of the interactions of quantum spacetime. By measuring the geometric properties on the corresponding quantum tetrahedra and simulate their interactions, our experiment serves as the basic module that represents the Feynman diagram vertex in the spin-network formulation of quantum spacetime.


\end{abstract}

\maketitle

\noindent
A quantum theory of gravity is one of the most fundamental questions of modern physics. Quantum gravity (QG) aims at incorporating the Einstein gravity with the principles of quantum mechanics, such that our understanding of gravity can be extended to the ultimate fundamental regime---the {\it Planck scale} $1.22 \times10^{19}$GeV \cite{kiefer2012quantum,book1,ashtekar,smolin2002three,Nicolai:2013sz}. At the Planck level, the Einstein gravity and hence the continuum spacetime break down, and what replaces these classical concepts is a quantum spacetime. Current approaches to quantum spacetime include string theory \cite{polchinski1998string}, loop quantum gravity (LQG) \cite{book}, twistor theory \cite{penrose1986spinors}, group field theory \cite{Freidel:2005qe}, dynamical triangulation \cite{Loll1998}, and Asymptotic safety \cite{Niedermaier2006}, etc. These approaches relate to a common framework of describing quantum spacetime, namely spin-networks, which is an important, non-perturbative tool of studying quantum spacetime.

\begin{figure}[!ht]
\includegraphics[width= \columnwidth]{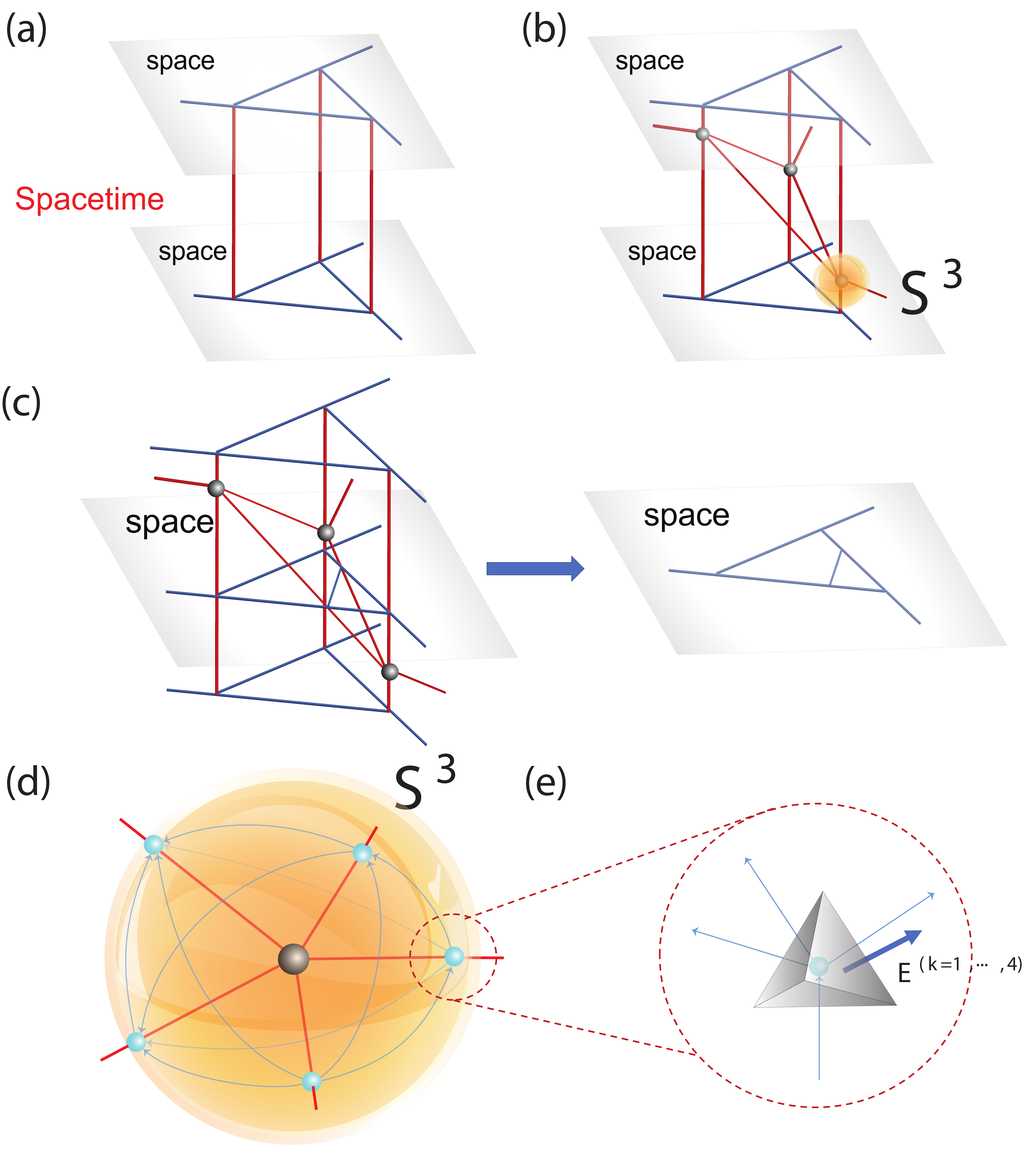}
\caption{(a) A static 4-dimensional quantum spacetime from evolving the spin-network. (b) A dynamical quantum spacetime with a number of vertices(in black) by intersecting intersecting world-sheets coloured by half-integer spins. (c) The intersection with an intermediate spatial slice gives an intermediate spin-network state, which is different from the initial state. The new link in the intermediate spin-network are the intersection between a world-sheet bounded by 3 vertices (in black) and the intermediate spatial slice. This demonstrates the dynamics given by the vertices. (d) The local structure of a vertex from (b) by considering a $3$-sphere  $S^3$ enclosing the vertex. Intersections between the world-sheets and $S^3$ give a spin-network (in blue, color online). The geometry is made by gluing $5$ (e) quantum geometrical tetrahedra. Each node of the spin-network associates with a quantum tetrahedron. Each face of a tetrahedron is dual to a link. Connecting $2$ nodes by a link in the spin-network corresponds to gluing $2$ tetrahedra through the face dual to the link.}
\label{sample}
\end{figure}

A spin-network is a graph whose (oriented) links and nodes are colored by half-integer spin labels (FIG.\ref{sample}(d)). Spin-networks are invented by Penrose, motivated by the twistor theory\cite{penroseSN}, then later on have been widely applied in LQG as the natural basis states in the Hilbert space of LQG\cite{Rovelli:1995ac,Rovelli1988,Ashtekar:1993wf,Bruegmann:1992gp,review1,review,Major:1999md}. Spin-networks also set up a framework for group field theories, which relate to dynamical triangulation and asymptotic safety. Some recent results exhibit the interesting relation between spin-networks and tensor networks in the anti de-Sitter/conformal field theory (AdS/CFT) correspondence originated from string theory\cite{hanhung,Singh:2017tet,Chirco:2017vhs}. Spin-networks have also been applied to gauge theories\cite{Baez1994,Baez:1994hx,Oeckl:2001wm,gambini2000loops} and related to topological orders in condense matter theories\cite{Levin:2004js,Konopka:2006hu,2011arXiv1106.6033K}.
There are extensive applications of spin-networks to topological invariants of manifolds of 3 and 4 dimensions, e.g., \cite{kauffman1993quantum,Turaev:1992hq,roland1,Crane:1994ji}.

We focus on $(3+1)$-dimensional quantum spacetime, in which case spin-networks are the quantum states of 3d Riemann geometries of the space (at the Planck scale), as the boundary data of quantum spacetime. As a profound prediction made by LQG, geometrical quantities, e.g. lengths, areas, and volumes, are quantized as operators on the Hilbert space of spin-network states, and have discrete eigenvalues \cite{Rovelli1995,ALarea,ALvolume,Bianchi:2008es,Thiemann:1996at,Ma:2010fy}. Quantum geometries at the Planck scale are fundamentally discrete, represented by spin-networks consisting of a number of 4-valent ($n$-valent) nodes. To be seen shortly, in a spin-network state, each $4$-valent node carries an invariant tensor of $SU(2)$, which depicts a quantum tetrahedron geometry (FIG.\ref{sample}(e)) \cite{Barbieri:1997ks,Baez:1999tk,shape,Rovelli:2006fw,CF,Bianchi:2011ub}. The $SU(2)$ invariance and the geometrical interpretation are consequences from the local Lorentz invariance in general relativity.

A quantum spacetime is a ``network'' in $3+1$ dimensions, consisting of a number of 2-dimensional world-sheets (surfaces) and their intersections, and the world-sheets are colored by half-interger spins. By the same token as the time evolution of a space builds up a classical spacetime, the time evolution of a spin-network forms a quantum spacetime \cite{rovelli2014covariant,Perez2012}. An example of a static quantum spacetime, where the spin-network does not evolve, is shown in FIG.\ref{sample}(a). In a quantum spacetime, each $1-$d spin-network link evolves to a $(1+1)$-d world sheet; hence the half-integer spin on the spin-network link can extend to the world-sheet. Dynamical quantum spacetimes (FIG.\ref{sample}(b)) are made by adding world-sheets (colored by spins) and their intersections, which creates a number of vertices. Vertices represent the local dynamics (interactions) of quantum geometry. Each vertex leads to a transition that changes the spin-network (FIG.\ref{sample}(c)). Quantum spacetimes made by intersecting world-sheets colored by half-integers are also called a \emph{spinfoam}. Similar to Feynman diagrams, quantum spacetimes associate transition amplitudes between initial and final spin-networks, called \emph{spinfoam amplitudes} \cite{Ooguri:1992eb,BC,EPRL,FK,KKL,NP,HT,LS,DFLS,generalize,HHKR}\footnote{Amplitudes of quantum spacetime are covariant by construction, independent of the choice of time direction. }. A spinfoam amplitude of a quantum spacetime is determined by the \emph{vertex amplitudes} locally associated to the intersection vertices in the quantum spacetime (FIG.\ref{sample}(d) and (e)). Quantum spacetimes and spinfoam amplitudes are a consistent and promising approach to QG \cite{propagator,semiclassical,Barrett:2009as,CFsemiclassical,HZ,propagator2,propagator3,claudio1,Alesci:2009ys,Rovelli:2010vv,HHKR,lowE,Han:2017xwo}.

In this work, we demonstrate quantum geometries of space and spacetime on a quantum simulator that simulates spin-networks and the building blocks of spinfoam amplitudes in $4$ dimensions. Using $4$-qubit quantum registers in the nuclear magnetic resonance (NMR) system, we create quantum tetrahedra and subsequently measure their quantum geometrical properties. Using the quantum tetrahedra in NMR, we simulate vertex amplitudes, which display the local dynamics of the corresponding quantum spacetime.
As quantum tetrahedra and vertex amplitudes serve as building blocks of large quantum spacetimes, our experiment opens up a new and practical way of studying quantum spacetimes and QG at large.



\underline{\emph{Quantum tetrahedron}:} Given a spin-network defined on an oriented graph $\G$. Each link $l$ is oriented and carries a half-integer $j_l$---an irreducible representation of $SU(2)$---that labels the ($2j_l+1$)-dimensional Hilbert space $\ch_{j_l}$ on the link labeled by $j_l$. Each $n$-valent vertex carries an invariant tensor $|i_n\rangle$ in the tensor representation $\otimes_l \ch_{j_l}$, i.e. $|i_n\rangle\in \inv{\otimes_l \ch_{j_l}},$ where $l$ labels the links incident (assumed all outgoing) at the vertex. On an ingoing link $l$,  $\ch_{j_l}$ is replaced by the dual $\ch_{j_l}^*$. A spin-network state is written as a triple $|\G,j_l,i_n\rangle$, defined by a tensor product of the invariant tensors at all nodes
\be
|\G,j_l,i_n\rangle:=\otimes_{n}|i_n\rangle\label{SN},
\ee
where spin labels of $|i_n\rangle$ are implicit. The $SU(2)$ invariance of $|i_n\rangle$ (the quantum constraint Eq.\Ref{closure}) is the gauge symmetry in QG, as the remanent from restricting the local Lorentz symmetry in a spatial slice \cite{book,review,review1}. All spin-networks with arbitrary $\G,j_l,i_n$ define an orthonormal basis in the Hilbert space of LQG.

Spin-network states Eq.\Ref{SN} are built by the  \emph{tensor product} of $|i_n\rangle$ at all nodes. Thus, simulating a spin-network with $m$ nodes, $\otimes_{n=1}^m|i_n\rangle$, only amounts to producing $m$ invariant tensors $|i_1\rangle,\cdots,|i_m\rangle$ in the experiment. It then suffices to simulate $|i_n\rangle$.

The rank $N$ of $|i_n\rangle$ coincides with the valence of the node $n$. In this letter, we mainly focus on $N=4$, which is of the most importance\footnote{The tetrahedron geometry with $N=4$ is the simplicial building block for arbitrary geometries in 3 dimensions.}. The $SU(2)$ invariance of a rank-$4$ $|i_n\rangle$ implies
\be
\lt(\hat{\bf J}^{(1)}+\hat{\bf J}^{(2)}+\hat{\bf J}^{(3)}+\hat{\bf J}^{(4)}\rt)|i_n\rangle=0.\label{closure}
\ee
Here, $\hat{\bf J}^{(k)}=(\hat{J}_x^{(k)},\hat{J}_y^{(k)},\hat{J}_z^{(k)})$ are the angular momentum operators on the Hilbert space $\ch_{j_k}$ carried by the $k$-th link of the four links meeting at the vertex. These operators satisfy $\hat{\bf J}^{(k)}\times\hat{\bf J}^{(k)}=i\, \hat{\bf J}^{(k)}$, where $\times$ is the vector product, and $[\hat{\bf J}^{(m)},\hat{\bf J}^{(k)}]=0$ if $m\neq k$. Interestingly, Eq.\Ref{closure} leads to a geometrical interpretation of invariant tensors and spin-networks.

On the other hand, the classical geometry of a tetrahedron in a 3d Euclidean space gives $4$ oriented areas ${\bf E}^{(k=1,\cdots,4)}=(E_x^{(k)},E_y^{(k)},E_z^{(k)})$, where $|{\bf E}^{(k)}|$ is the area of the $k$-th face, and ${\bf E}^{(k)}/|{\bf E}^{(k)}|$ is the unit vector normal to the face. The four faces of a tetrahedron form a closed surface, namely,
\be
{\bf E}^{(1)}+{\bf E}^{(2)}+{\bf E}^{(3)}+{\bf E}^{(4)}=0. \label{cclosure}
\ee
Conversely, the data ${\bf E}^{(k=1,\cdots,4)}$ subject to constraint \Ref{cclosure} uniquely determine the (Euclidean) tetrahedron geometry \cite{Minkowski}. Euclidean tetrahedra are the fundamental building blocks of arbitrary curved 3d geometries, since any geometry can be triangulated and approximated by a large number of Euclidean tetrahedra.

\begin{figure*}[!htbp]
\includegraphics[width=2 \columnwidth]{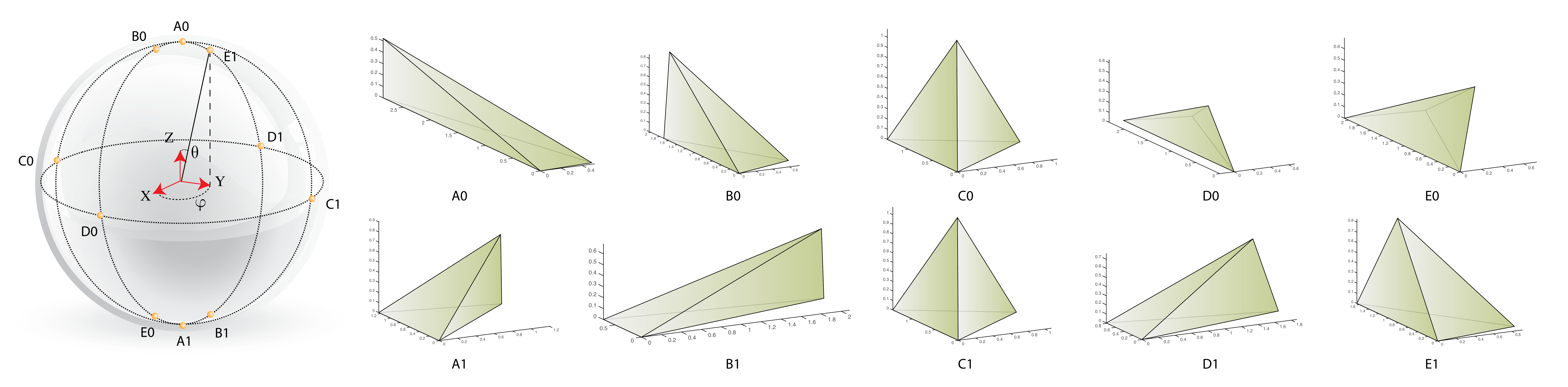}
\caption{Experimentally prepared states on the Bloch sphere and their corresponding classical tetrahedra. The states take the form $\cos{\frac{\theta}{2}}|0\rangle_{L}+e^{i\phi}\sin{\frac{\theta}{2}}|1\rangle_{L}$, where the north and south poles of the Bloch sphere are typically chosen to correspond to the standard basis vectors $|0\rangle_{L}$ and $|1\rangle_{L}$. The $10$ experimental prepared quantum states $\mathrm{Ai, Bi, Ci, Di, Ei\  (i=0,1)}$ and their corresponding tetrahedra are are shown on the right.}
\label{exp_state}
\end{figure*}

Comparing Eqs. (\ref{closure}) and \Ref{cclosure} suggests the quantization of tetrahedron geometries. That is, $\hat{\bf J}^{(k)}$ is the quantum version of ${\bf E}^{(k)}$, so is Eq. \eqref{cclosure} to Eq. \eqref{closure}. Precisely, we have
\be\label{eq:areaOperator}
\hat{\bf E}^{(k)}=8\pi\ell_P^2\hat{\bf J}^{(k)},
\ee
where $a,b,c=x,y,x$, $\eps_{abc}$ is the Levi-Civita symbol, $G_N$ is the Newton's constant, and $\ell_P$ is the Planck length.
More detailed physical account for this quantization is left to Appendix \ref{appd:lqgQuantization}.

Quantum gravity identifies quantum-tetrahedron geometries with a system of quantum angular momentums subject to Eq. \Ref{closure}. This identification enables us to simulate quantum geometries with qubits.  We focus on the situation with all spins $j_k=1/2$ ($\ch_{j=1/2}\simeq \C^2$) and simulate the quantum tetrahedra with 4-qubit tensor states in $\ch_{j=1/2}^{\otimes 4}$. Invariant tensors of 4 qubits spans a 2-dimensional subspace $\inv{\ch_{j=1/2}^{\otimes 4}}$ (See Appendix~\ref{appd:dimension} for details.). Each invariant tensor $|i\rangle $ turns out to reconstruct a quantum-tetrahedron geometry. Tetrahedron geometries are now encoded in a quantum Hilbert space of invariant tensors.

\underline{\emph{Quantum spacetime atom}:} Let's come back to the spin-network state $\otimes_{n=1}^5 |i_n\rangle$ in FIG.\ref{sample}(d) made by $5$ quantum tetrahedra. This state is the boundary state of a vertex in a quantum spacetime. Indeed, given a $4$d quantum spacetime shown in FIG.\ref{sample}(b), we consider a $3$-sphere enclosing a portion of the quantum spacetime surrounding a vertex. The boundary of the enclosed quantum spacetime is precisely a spin-network (see FIG.\ref{sample}(d)). Large quantum spacetimes with many vertices can be obtained by gluing such portions FIG.\ref{sample}(b). Such a portion of FIG.\ref{sample}(b) is an atom of quantum spacetimes.

An atom of quantum spacetimes associates with a vertex amplitude, which is an evaluation of the spin-network $\otimes_{n=1}^5 |i_n\rangle$. The evaluation maps a spin-network to a number, or more precisely a function of $5$ invariant tensors. Let's consider $5$ quantum tetrahedra made by 4-qubit invariant tensors $|i_n\rangle$ ($n=1,\cdots,5$), each of which associates with a node in the spin-network (blue in FIG.\ref{sample}(d)). Each Hilbert space $\ch_{j=1/2}$ for tensors $|i_n\rangle\in \inv{\ch_{j=1/2}^{\otimes 4}}$ associates with a link in the spin-network. We consider the following evaluation of $\otimes_{n=1}^5 |i_n\rangle$ by picking up the $2$-qubit maximally entangled state $|\epsilon_l\rangle=(|01\rangle-|10\rangle)/\sqrt{2}$ for each link $l$, where the two qubits associate respectively with the end points of $l$. The evaluation is given by the inner product
\be
\bigotimes_{l=1}^{10}\langle \epsilon_l |\ \bigotimes_{n=1}^5 |\,i_n\,\rangle=A(i_1,\cdots,i_5).
\label{vetxamp}
\ee
The inner product above takes place at the end points of each $l$, between a qubit in $|\epsilon_l\rangle$ and the other in $|i_n\rangle$. The resultant $A(i_1,\cdots,i_5)$ is the vertex amplitude of the quantum spacetime at the Planck level in Ooguri's model\cite{Ooguri:1992eb}, where the spins on the world-sheets are all $1/2$. Ooguri's model defines a topological invariant of $4$-manifolds. Vertex amplitudes in Ooguri's model relate to the classical action of gravity when the spins are large\cite{Barrett:2009as}.

The spin-network $\otimes_{n=1}^5 |i_n\rangle$ shows the (quantum) gluing of $5$ tetrahedra to form a closed $S^3$ in FIG. \ref{sample}(d). Each link in the spin-network corresponds to gluing a pair of faces of 2 different tetrahedra. Such gluing does not require the faces being glued to match in shape because of quantum fluctuations but to match in their quantum area $\Ar_k=8\pi\ell_P^2\sqrt{3/4}$. Quantum geometries on $S^3$ are unsmooth. The vertex amplitude $A(i_1,\cdots,i_5)$ is the transition amplitude from $m$ to $5-m$ quantum tetrahedra $(m<5)$, or covariantly, the interaction amplitude of $5$ quantum tetrahedra. Such amplitudes describe the local dynamics of QG in the $4$d quantum spacetime enclosed by the $S^3$.

\underline{\emph{Experimental design and implementation}}
Reconstructing quantum tetrahedra makes use of various geometrical operators on $\inv{\ch_{j=1/2}^{\otimes 4}}$. Using the quantization \eqref{eq:areaOperator}, the quantum area of the $k$-th face is diagonalized\cite{Rovelli1995,ALarea} as
\be
\widehat{\Ar}_k|i\rangle&=&\sqrt{\hat{\bf E}^{(k)}\cdot \hat{\bf E}^{(k)}}\ |i\rangle=8\pi\ell_P^2\sqrt{3/4}\ |i\rangle.\label{qarea}
\ee
The expectation value of an area operator in an invariant tensor $|i\rangle $ is $\langle i|\widehat{\Ar}_k|i\rangle=8\pi\ell_P^2\sqrt{3/4}$. In addition, dihedral angles $\theta_{km}$ between the $k$-th and $m$-th faces are quantized accordingly\cite{Rovelli:2006fw}
as\be
\widehat{\cos\theta}_{km}=\frac{\hat{\bf E}^{(k)}\cdot \hat{\bf E}^{(m)}}{\sqrt{\hat{\bf E}^{(k)}\cdot \hat{\bf E}^{(k)}}\sqrt{\hat{\bf E}^{(m)}\cdot \hat{\bf E}^{(m)}}}
=\frac{4}{3}\ {\hat{\bf J}^{(k)}\cdot\hat{\bf J}^{(m)}}.  \label{dian}
\ee
Because of Eq. \Ref{closure}, there are only two independent expectation values of $\widehat{\cos\theta}_{km}$, say, $\langle i|\widehat{\cos\theta}_{12}|i\rangle$ and $\langle i|\widehat{\cos\theta}_{13}|i\rangle$. In an $|i\rangle$, the expectation values of the four areas and two dihedral-angle operators uniquely determine a geometrical tetrahedron. (See details in Appendix~\ref{appd:freedom}.)
Since $\inv{\ch_{j=1/2}^{\otimes 4}}$ is $2$-dimensional, it can be presented as a Bloch sphere. Any point $(\theta,\phi)$ on the Bloch sphere uniquely reconstructs a quantum tetrahedron geometry as shown in FIG. \ref{exp_state}, whose area of each face is $8\pi\ell_P^2\sqrt{3/4} $ and the mean value of $2$ independent dihedral-angles can be  calculated  by Appendix ~\ref{appd:angle}.

%

The experimental target quantum tetrahedron states are labeled by 10 orange balls on the Bloch sphere as shown in FIG. \ref{exp_state}, whose spherical coordinates are listed in Table.\ref{state} in Appendix \ref{app:exppart}.

All experiments were carried out on a 700MHz DRX Bruker spectrometer, at the temperature of 298K. The Crotonic Acid molecule, whose details can be found in Appendix \ref{app:exppart}, works as our four-qubit quantum system. To prepare the fundamental building blocks---quantum tetrahedra and simulate the local dynamics of quantum spacetimes, we divide the whole experiment into three parts as follows.

\textit{States Preparation---} The NMR experiment always begin with the thermal equilibrium state. First, we initialized the whole system to a pseudo-pure state (PPS) with the fidelity over 99\%. More details about PPS are put into Appendix \ref{app:exppart}. Then, the system were driven into each of the states representing the target tetrahedra, as shown in Fig. \ref{exp_state}, respectively. In this step, we denote the experimentally prepared state as $\rho_{i}^{tetra}$, where $i=A0,A1...E0,E1$. There are ten pulses bridging the PPS and the ten quantum tetrahedra. Those pulses were realized by the gradient ascent pulse engineering (GRAPE) optimizations, with the length of $20$ms.

\textit{Measure Geometry---}Generally speaking,  a tetrahedron can be uniquely determined by six independent constrictions. Since the identity part generates no signal in our NMR system, the area operators defined in Eq. \eqref{qarea} are unmeasurable. In the experiment,  we stress on dihedral angles $\widehat{\cos\theta}_{km}$ defined in Eq. \eqref{dian}, where $k\neq m$ and $k,m=1...4$. These $\widehat{\cos\theta}_{km}$ can take a form in terms of Pauli matrices: $(\sigma_{x}^{k}\sigma_{x}^{m}+\sigma_{y}^{k}\sigma_{y}^{m}+\sigma_{z}^{k}\sigma_{z}^{m})/6$. The observables such as $trace(\sigma_{x}^{k}\sigma_{x}^{m},\rho_{i}^{tetra}) (i=A_0,A_1...E_0,E_1)$ can be easily measured by adding an observable pulse after the state preparation, which function as single-qubit rotation and was optimized with a $1$ms GRAPE pulse.

\begin{figure}[htpp]
\includegraphics[width= 8cm]{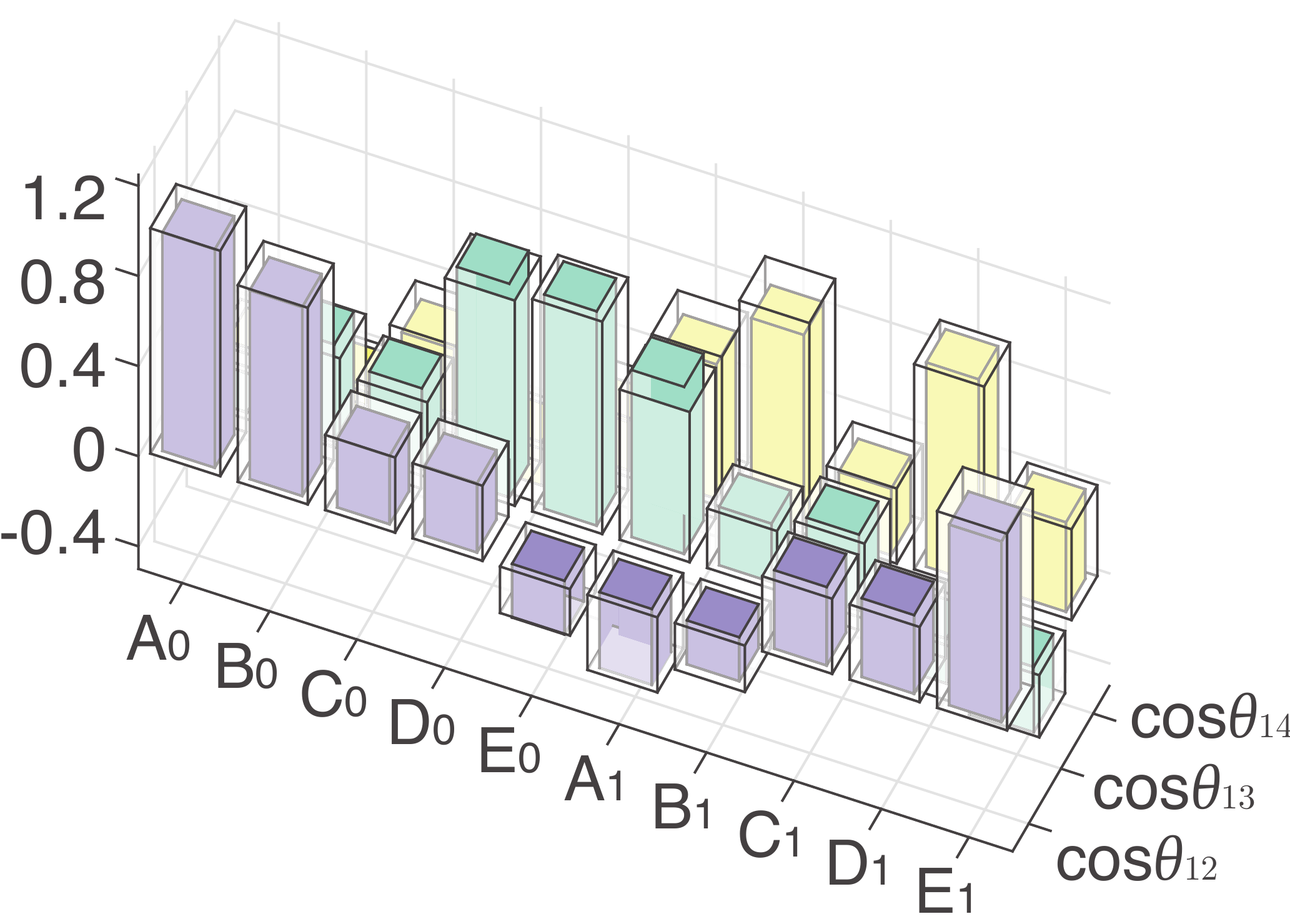}
\caption{The cosine value of the three dihedral angles for $10$ prepared quantum tetrahedron state: vertical axis represents the cosine value of the dihedral angles between the bottom face and the others, while the coordinates of the horizontal plane indicate the dihedral angle and $10$ prepared states, respectively.  Besides, the transparent columns represent the theoretical values, while the coloured ones represent the experimental results.}
\label{geo}
\end{figure}

We present the measured geometry properties via a $3$-dimensional histogram (Fig. \ref{geo}), whose vertical axis represents the cosine value of the dihedral angles between the bottom face and the others. In the figure, the transparent columns represent the theoretical values, while the coloured ones represent the experimental results. The maximum difference between experiment and the theory is within $0.08$. From the figure, It can be said that our experimental prepared states matches the building blocks---quantum tetrahedra successfully.

Since those geometrical operators do not commute, they have quantum fluctuations. There are three independent quadratic fluctuations of dihedral angles $\Delta_{km}:=\lt(\widehat{\cos\theta}_{km}-\langle i|\widehat{\cos\theta}_{km}|i\rangle\rt)^2$, say, $(k,m)=(1,2),(1,3),(1,4)$. In this paper we shall add these three $ \Delta_{km}$ to be the total quantum fluctuation of the quantum tetrahedron (see Appendix ~\ref{appd:angle})
\begin{eqnarray}
\Delta=\Delta_{12}+\Delta_{13}+\Delta_{14}=\frac{2}{3}  +\frac{8}{3}\cos^2\frac{\theta}{2}\sin^2\frac{\theta}{2}(1-\cos^2\phi).
\label{flua}
\end{eqnarray}
The experimentally prepared states are all in the minimal fluctuation of area since the second term of Eq. \eqref{flua} always equals to $0$. The fluctuation defined above are all $2/3,$ while the experimentally measured values are listed in Table. \ref{state} of  Appendix \ref{app:exppart}.
\begin{table*}[htpp]
\caption{The regular tetrahedra $|i_{n}\rangle(n=1...4)$  and $|i_{5}\rangle$ in  Eq. \ref{eq:invariant} are replaced with the experiment states. We list the real and imaginary part of the amplitude:}
\begin{center}
\begin{tabular}{|cc|c|c|c|c|c|c|c|c|c|c|}
\hline
& &$A_0$ & $B_0$ & $C_0$ & $D_0$ & $E_0$&$A_1$ & $B_1$ & $C_1$ & $D_1$ & $E_1$ \\
\hline
\hline
\multirow{2}{*}{Re($10^{-5}$)}&theory & -13.5635 &-20.1590 &   0.0000 & -26.2024&  -26.5339 &  23.4924 &  18.1513  &-27.1270 &   7.0208& -5.6401 \\
\cline{2-12}
&experiment & -12.74 & -19.89  & 0.01 & -24.59&  -25.72 & -22.16 &  18.73 & -25.48&    4.32&   -3.84 \\
\hline
\hline
\multirow{2}{*}{Im($10^{-5}$)}&theory&-23.4923 & -18.1514  &  0.0000 &  -7.0210 &   5.6400  &-13.5634 & -20.1591  &-46.9848&  -26.2024 & -26.5339 \\
\cline{2-12}
&experiment & -23.67&  -17.78 &   0.05  & -7.98  &  6.63  & 13.16&  -18.10 & -44.14 & -25.62 & -25.86 \\
\hline
\end{tabular}
\end{center}
\label{tableamp}
\end{table*}%


Those quantum fluctuations are large because quantum tetrahedra are simulated by qubits with $j=1/2$. These tetrahedra are of Planck size ($\Ar\sim \ell_P^2$) and typically appear in quantum spacetime near the big bang or a black hole singularity \cite{Han:2016fgh}. Invariant tensors with spins $j\gg 1$ exhibit tetrahedron geometries with small quantum fluctuations\cite{LS,Rovelli:2006fw}.

\emph{Simulate the Amplitudes---}As the vertex amplitude stated in Eq. \eqref{vetxamp} can describe the the local dynamics of QG in the 4d quantum spacetime, to obtain such amplitudes, we need to calculate the inner products between different quantum tetrahedron states.  We do not implement the real dynamics of the spin-foam consisting of five tetrahedra, which would need a $20$-qubit quantum register. Alternatively, a full tomography follows our state preparation to obtain the information of quantum tetrahedron states. The fidelities between the experimentally prepared quantum tetrahedron states and the theoretical ones were also calculated. They are all above 95\% and the details can be seen in Appendix \ref{app:exppart}.

\begin{figure}[!ht]
\includegraphics[width= 8cm]{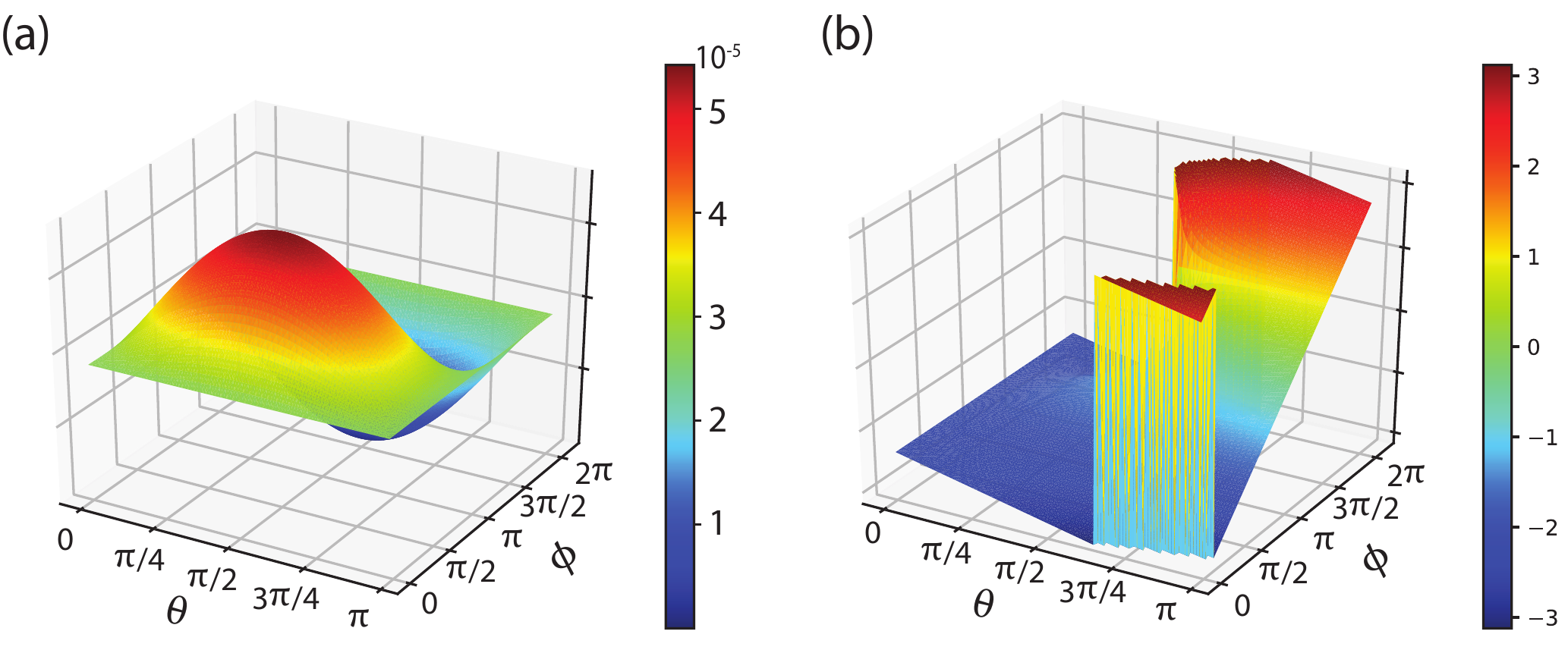}
\caption{Simulation results of the vertex amplitudes in Ooguri's model: We fix the regular tetrahedra $|i_{n}\rangle(n=1...4)$  and alter the $|i_{5}\rangle$ by varying $\theta$ and $\phi$, as Eq. \eqref{eq:invariant} shows in Appendix. (a) is the amplitude of Eq. \ref{vetxamp} while (b) discribe the information of its phase.}
\label{amp}
\end{figure}

To present the consequences more intuitively, the $|i_{n}\rangle(n=1...4)\rangle$ in Eq. \eqref{vetxamp} are fixed as regular quantum tetrahedra, while the spherical coordinates $\theta$ and $\phi$ of $|i_{5}\rangle$ varied smoothly. Fig. \ref{amp}(a) and \ref{amp}(b) show the simulation results, with the value of the amplitude and phase, respectively. Mixed states  are  inevitably introduced to the experiment since inevitably experimental error. To calculate the inner products in the vertex amplitude formula in Eq. \eqref{vetxamp}, we purified the measured density matrices, using the method of maximal likelihood. The comparison between the experiment and the numeric simulation are listed in Table. \ref{tableamp}.


\underline{\emph{Conclusion}}---Our experiment is the initial endeavour to simulate quantum tetrahedra---the building blocks of spin-networks and hence of quantum spacetimes at the Planck level. By creating ten different quantum tetrahedra on our NMR quantum simulator, we measure their dihedral-angles and simulate the vertex amplitudes. As the first step towards exploring spin-networks using a quantum simulator, our work provides valid experimental demonstrations about studying quantum spacetimes to date.

\begin{acknowledgements}
This research was supported by CIFAR, NSERC and Industry of Canada. K.L. and G.L. acknowledge National Natural Science Foundation of China under Grants No. 11175094 and No. 91221205. Y.L. acknowledges support from Chinese Ministry of Education under grants No.20173080024. MH acknowledges support from the US National Science Foundation through grant PHY-1602867, and startup grant at Florida Atlantic University, USA. YW thanks the startup grant offered by the Fudan University and the hospitality of IQC and PI during his visit, where this work was partially conducted.
YW is also supported by the Shanghai Pujiang Program No. KBH1512328. D. L. is supported by Guangdong Innovative and Entrepreneurial Research Team Program (No. 2016ZT06D348).\end{acknowledgements}

\appendix
\section{LQG Quantization}\label{appd:lqgQuantization}
The relation is indeed precise by the quantization of gravity with Ashtekar's new variables \cite{Ashtekar:1986yd,barbero}. Einstein gravity identifies gravity with Riemannian geometry; hence, dynamical variables of gravity relates to geometrical variables such as ${\bf E}^{(k)}$. The Poisson bracket of gravity variables endows the following Poisson bracket to ${\bf E}^{(k)}$ \cite{Ashtekar:1998ak,book}
\be
\lt\{E_a^{(m)},E_b^{(k)}\rt\}=8\pi G_N\sum_{c}\eps_{abc}E_c^{(k)}\,\delta^{mk}.
\ee
 The quantization promotes ${\bf E}^{(k)}$ to operators $\hat{\bf E}^{(k)}$. Interestingly $[\ ,\ ]=i\hbar\{\ ,\ \}$ gives precisely the commutation relation of the angular momentum operators $\hat{\bf J}^{(k)}$'s in quantum mechanics (the identification Eq. \eqref{eq:areaOperator}). Each $\hat{\bf J}^{(k)}$ acts on the irreducible representation $\ch_{j_k}$ of $SU(2)$ labelled by a spin $j_k\in \Z/2$. The Hilbert space of a quantum tetrahedron is the space of rank-$4$ invariant tensors $\inv{\otimes_{k=1}^4 \ch_{j_k}}$, as solutions of the quantum constraint Eq. \eqref{closure}.

\section{Invariant Subspace and Logic Bit}\label{appd:dimension}
When considering a system with more than one subsystem, in which angular
momentum is a good quantum number for both the individual subsystems and
the whole system, we can represent system in different basis.
For instance, a system with two particles, we have two different representations
\[
\ket{j_1\,m_1}\ket{j_2\,m_2},
\]
where $m_i \in \{-j_i,-j_i+1,\cdots,j_i\}$, and
\[
\ket{j_1\,j_2\,J_{1\,2},M_{1\,2}},
\]
where $J_{1\,2} \in \{|j_1-j_2|,|j_1-j_2|+1,\cdots,j_1+j_2\} $ (known as the triangle condition),
$M_{1\,2} \in \{-J_{1\,2},-J_{1\,2}+1,\cdots,J_{1\,2}\}$ and $M_{1\,2}=m_1+m_2$. $J_{1\,2}$ and $M_{1\,2}$ together describe the angular momentum of the whole space. These two representations are related by a unitary transformation
\begin{eqnarray*}
& &\ket{j_1\,j_2\,J_{1\,2},M_{1\,2}}
\sum_{m_1=-j_1}^{j_1}\sum_{m_2=-j_2}^{j_2}C^{\, j_1\,\, j_2}_{m_1\, m_2\, J_{1\,2},\, M_{1\,2}}\ket{j_1\,m_1}\ket{j_2\,m_2}.
\end{eqnarray*}
Here, $C^{\, j_1\,\, j_2}_{m_1\, m_2\, J_{1\,2},\, M_{1\,2}}$ are the Clebsch-Gordan coefficients, which can all be chosen to be real numbers.

When we consider a system with four particles, whose spins are $j_1$, $j_2$, $j_3$ and $j_4$ respectively, we
can couple the particles $1$ and $2$ to get an intermediate angular momentum,
say $J_{1\,2}$. At the same time, we couple the particles $3$ and $4$ to get $J_{3\,4}$.
Finally, we
choose possible values of
among all $J_{1\,2}$ and $J_{3\,4}$ to get the total angular momentum $J$.

\begin{displaymath}
\xymatrix{
  j_1\ar[dr]     &          & j_2\ar[dl] & & j_3\ar[dr] &          & j_4\ar[dl]    \\
                 & J_{1\,2}\ar[drr] &            & &            & J_{3\,4}\ar[dll] &          \\
                 &          &            &J &            &          &
          }
\end{displaymath}

%
%
%

Although the initial spins $j_1$, $j_2$, $j_3$ and
$j_4$ as well as the final $J$ are fixed, the intermediate
angular momenta can be arbitrary, as long as the triangle condition
holds in each step.

When $j_1=j_2=j_3=j_4=\frac{1}{2}$ and the final $J=0$
(i.e. the $4$-qubit invariant tensor situation),
the triangular condition requires $J_{1\,2}=J_{3\,4}$ to meet $J=0$, but $J_{1\,2}$ can be either $0$ or $1$. Obviously, the
dimension of the invariant subspace is $2$. A general
invariant $4$-qubit tensor reads
\begin{eqnarray}
\ket{\psi_4}&=&\sum_{J_{1\,2}=0,1}\alpha(J_{1\,2})\ket{\phi_{J_{1\,2}}}\notag\\
&=&\frac{\alpha(0)}{2}\bigg(\ket{0 1}-\ket{1 0}\bigg)\bigg(\ket{0 1}-\ket{1 0}\bigg)\notag\\
& &+\frac{\alpha(1)}{\sqrt{3}}\left[\ket{1 1 0 0}+\ket{0 0 1 1} -\frac{1}{2}\big(\ket{0 1}+\ket{1 0}\big)\big(\ket{0 1}+\ket{1 0}\big)\right]\notag\\
&=&\cos \frac{\theta}{2} \ket{0_L}+e^{i\phi}\sin \frac{\theta}{2}\ket{1_L},\label{eq:invariant}
\end{eqnarray}
where
\begin{eqnarray*}
\ket{0_L}&=&\frac{1}{2}\bigg(\ket{0 1}-\ket{1 0}\bigg)\bigg(\ket{0 1}-\ket{1 0}\bigg),\\
\ket{1_L}&=&\frac{1}{\sqrt{3}}\left[\ket{1 1 0 0}+\ket{0 0 1 1} -\frac{1}{2}\big(\ket{0 1}+\ket{1 0}\big)\big(\ket{0 1}+\ket{1 0}\big)\right],
\end{eqnarray*}
are the logical-bit representation of this subspace. As usual, $\theta$ and $\phi$ uniquely determine a state on the Bloch sphere.

\section{Freedom of Classical Tetrahedra}\label{appd:freedom}

A tetrahedron has $4$ faces and each possesses $3$ parameters. Two of the parameters describe the direction of the face and one parameter for the area. Therefore, given an arbitrary tetrahedron, we have $12$ parameters. Nevertheless, these arbitrary tetrahedra fall into different equivalent classes. In each of equivalence class, the tetrahedra transform into each other by translations and rotations in $3$ dimensions. This equivalence eliminates $6$ of the 12 parameters, leaving only $6$ independent parameters, which can be chosen to be the $4$ face areas and $2$ independent dihedral angles.

Once given the $4$ face areas, $A_1, A_2, A_3$ and $A_4$, and $2$ independent dihedral-angles, say, $\theta_{1,2}$, one can determine the tetrahedron in the following procedure:
\begin{enumerate}
\item Let vertex $A$ be the coordinate origin, vertex $B$ on $\{a,0,0\}$, vertex $C$ on $\{b,c,0\}$ and the last vertex $D$ on $\{d,e,f\}$, then label faces $ABC$, $ACD$, $ABD$ and $BCD$ as $1,2,3$ and $4$ respectively;
\item write down the $6$ constraints of the areas and dihedral angles;
\item Obtain the solution $\{a,b,c,d,e,f\}$ that determines the tetrahedron.
\end{enumerate}

\section{Mean Value and Quantum Fluctuation of Dihedral-Angles}\label{appd:angle}

For any tetrahedron, there are $6$ different dihedral angles $\theta_{ij}$. The operators $\widehat{\cos \theta_{ij}}$ are defined in Eq.~(\ref{dian}).
Due to the closure condition Eq.~(\ref{closure}), one can derive
\begin{eqnarray*}
&&\langle \widehat{\cos \theta_{12}} \rangle=\langle\widehat{\cos \theta_{34}}\rangle,\\
&&\langle\cos \widehat{\theta_{13}}\rangle=\langle\widehat{\cos \theta_{24}}\rangle,\\
&&\langle\cos \widehat{\theta_{14}}\rangle=\langle\cos \widehat{\theta_{23}}\rangle,\\
&&\langle\cos \widehat{\theta_{12}}\rangle + \langle\widehat{\cos \theta_{13}}\rangle + \langle\widehat{\cos \theta_{14}}\rangle=1.
\end{eqnarray*}
Thus, there are only $2$ independent such operators, and we shall take $\widehat{\cos \theta_{12}}$ and $\widehat{\cos \theta_{13}}$ without loss of generality.
The operator $\widehat{\cos\theta_{12}}$ is diagonal in the basis we use to describe the invariant subspace in Appendix ~\ref{appd:dimension}, which are the eigenstates of the operator. Define $\ket{0_L}=(0,1)^T$ and $\ket{1_L}=(1,0)^T$, one can easily check that
\begin{eqnarray*}
&&\widehat{\cos \theta_{12}}=\left(
\begin{tabular}{c c}
$-\frac{1}{3}$ & $0$\\
$0$  &  $1$
\end{tabular}
\right),\qquad
\widehat{\cos^2 \theta_{12}}=\left(
\begin{tabular}{c c}
$\frac{1}{9}$ & $0$\\
$0$  &  $1$
\end{tabular}
\right),\\
&&\widehat{\cos \theta_{13}}=\left(
\begin{tabular}{c c}
$\frac{2}{3}$ & $\frac{\sqrt{3}}{3}$\\
$\frac{\sqrt{3}}{3}$ & $0$
\end{tabular}
\right),\quad\,\,
\widehat{\cos^2 \theta_{13}}=\left(
\begin{tabular}{c c}
$\frac{7}{9}$ & $\frac{2\sqrt{3}}{9}$\\
$\frac{2\sqrt{3}}{9}$ & $\frac{1}{3}$
\end{tabular}
\right),\\
&&\widehat{\cos \theta_{14}}=\left(
\begin{tabular}{c c}
$\frac{2}{3}$ & $-\frac{\sqrt{3}}{3}$\\
$-\frac{\sqrt{3}}{3}$ & $0$
\end{tabular}
\right),\,
\widehat{\cos^2 \theta_{14}}=\left(
\begin{tabular}{c c}
$\frac{7}{9}$ & $-\frac{2\sqrt{3}}{9}$\\
$-\frac{2\sqrt{3}}{9}$ & $\frac{1}{3}$
\end{tabular}
\right),
\end{eqnarray*}
the mean value of $2$ independent dihedral-angles under the state $(\theta,\phi)$ on the Bloch sphere can be chosen as
\begin{eqnarray}
\langle \widehat{\cos \theta_{12}} \rangle&=&\cos^2\frac{\theta}{2}-\frac{1}{3}\sin^2\frac{\theta}{2},\\
\langle \widehat{\cos \theta_{13}} \rangle&=&\frac{2}{3}\sin^2\frac{\theta}{2}+\frac{2\sqrt{3}}{3}\cos\frac{\theta}{2}\sin\frac{\theta}{2}e^{i\phi}.
\end{eqnarray}

Thus the quantum fluctuation on the invariant tensor in Eq.~(\ref{eq:invariant}) reads
\begin{eqnarray}
\Delta=\Delta_{12}+\Delta_{13}+\Delta_{14}=\frac{2}{3}  +\frac{8}{3}\cos^2\frac{\theta}{2}\sin^2\frac{\theta}{2}(1-\cos^2\phi),
\end{eqnarray}
where $\Delta_{km}:=\lt(\widehat{\cos\theta}_{km}-\langle i|\widehat{\cos\theta}_{km}|i\rangle\rt)^2$.

\section{exp\_part}\label{app:exppart}
\begin{figure}[!ht]
\includegraphics[width= 8cm]{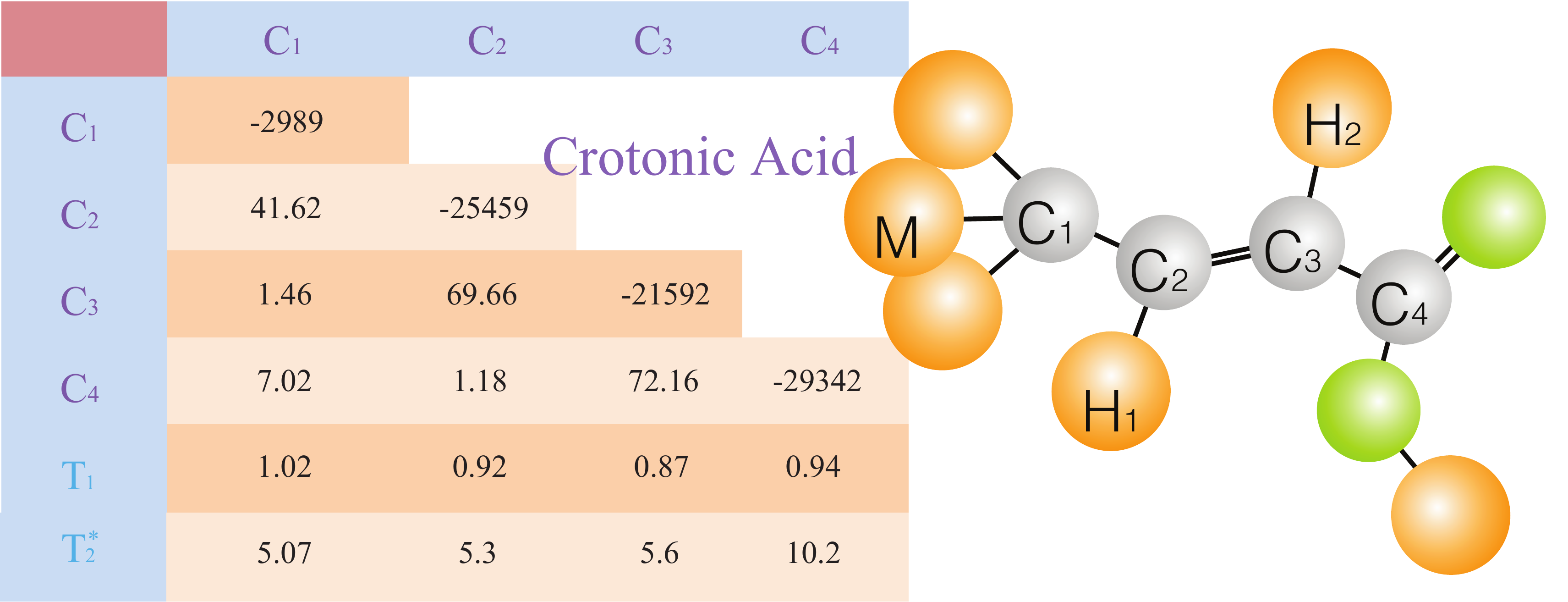}
\caption{Structure of Crotonic Acid molecule; The four $^{13}C$ nuclei are denoted as the four qubits and the table on the left presents the parameters constructing the internal Hamiltonian. Chemical shifts (Hz),  J-coupling strengths (Hz) and  and the relaxation times( T$_1$ and T$_2$) are listed in the diagonal part, off-diagonal elements and the bottom, respectively. All parameters were measured on a Bruker DRX 700 MHz spectrometer at room temperature. }
\label{molecule}
\end{figure}

\emph{Molecule---}All experiments are based on a Crotonic Acid molecule, dissolved in the d6-acetone, whose structure are depicted in Fig. \ref{molecule}. The internal Hamiltonian of the system under weak coupling approximation is
\begin{align}\label{Hamiltonian}
\mathcal{H}_{int}=\sum\limits_{j=1}^4 {\pi \nu _j } \sigma_z^j  + \sum\limits_{j < k,=1}^4 \frac{\pi}{2} J_{jk} \sigma_z^j \sigma_z^k,
\end{align}
where $\nu_j$ is the chemical shift of the \emph{j}th spin and $\emph{J}_{jk}$ is the spin-spin interaction($J$-coupling) strength between spins \emph{j} and \emph{k}.

\begin{figure}[!ht]
\includegraphics[width= 8cm]{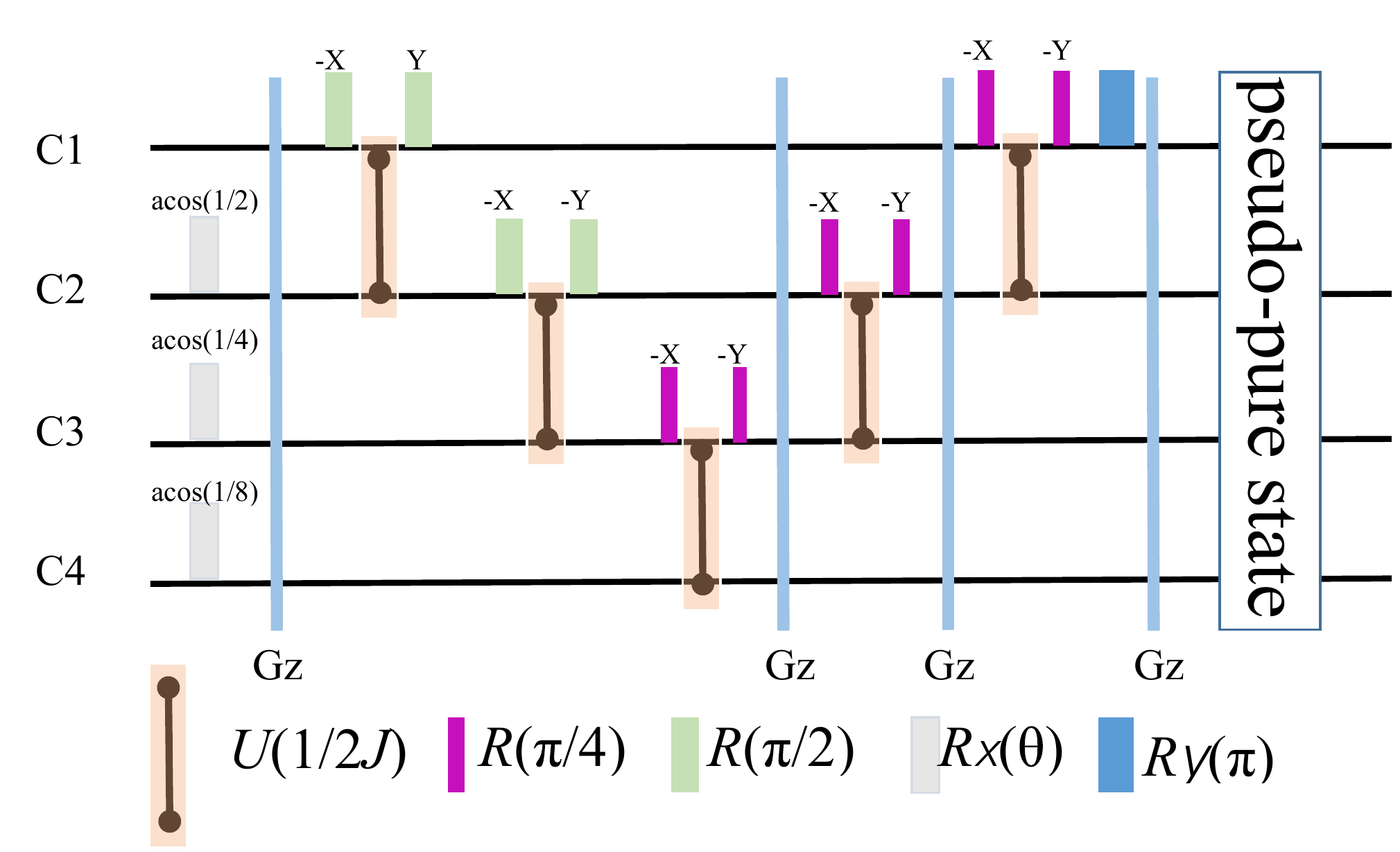}
\caption{Structure of Crotonic Acid molecule; The four $^{13}C$ nuclei are denoted as the four qubits and the table on the left presents the parameters constructing the internal Hamiltonian. Chemical shifts (Hz),  $J$-coupling strengths (Hz) and  and the relaxation times( $T_1$ and $T_2$) are listed in the diagonal part, off-diagonal elements, and the bottom, respectively. All parameters were measured on a Bruker DRX $700$ MHz spectrometer at room temperature.}
\label{pps_cir}
\end{figure}

\emph{Pseudo-pure state---}The four-qubit NMR system begins with the thermal equilibrium state $\rho_{eq}$:
\begin{equation}
\rho_{eq}=\frac{1-\epsilon}{16} \mathbb{I} + \epsilon ( \sigma_{z}^{1}+\sigma_{z}^{2}+\sigma_{z}^{3}+\sigma_{z}^{4}),
\end{equation}
 where $\epsilon \approx 10^{-5}$ describes the polarization when setting gyromagnetic ratio of $^{13}C$ to 1, and $ \mathbb{I}$ is a $16\times 16$ identity matrix. To create the pseudo-pure state
\begin{equation}
\rho_{0000}=\frac{1-\epsilon}{16} \mathbb{I}+ \epsilon |0000\rangle\langle0000|,
\end{equation}
we used the spatial average technique shown in Fig. \ref{pps_cir}, which includes four $z$-gradient fields. In between any two gradient fields, the free evolution was implemented by inserting $\pi$ pulses and all local operations were realized by $1$ms GRAPE pulses. Consequently, the fidelity of the experimentally prepared PPS is above 99\%. As the identity part does not influence the unitary operations or measurements in NMR experiments, the original density matrix of $\rho_{0000}$ can be replaced by the deviated one for simplicity. The state $\rho_{0000}=\epsilon |0000\rangle\langle0000|$ is taken as the referential state in our following experiments.

\begin{table*}[ht]
\caption{Parameters and Geometry fluctuation for the experimental prepared quantum tetrahedron states. The first two rows represent the spherical coordinates where $A_{i}\sim E_i(i=0,1)$ and $\theta$ ,$\phi$ are the labels depicted in Fig.\ref{exp_state}. Last two rows shows the experimental and theoretical value of the fluctuations of dihedral angles in quantum tetrahedron}.
\begin{center}
\begin{tabular}{|c|cccccccccc|}
\hline
&$A_0$ & $B_0$ & $C_0$ & $D_0$ & $E_0$&$A_1$ & $B_1$ & $C_1$ & $D_1$ & $E_1$ \\
\hline
\hline
$\theta$& 0 &$\pi/5$ & $\pi/2$ & $\pi/2$  & $4\pi/5$& $\pi$ & $4\pi/5$  & $\pi/2$ & $\pi/2$ & $\pi/5$  \\
\hline
$\phi$ &0 & 0 & $3\pi/2$  &0   & 0& $\pi$ & $\pi$  &  $\pi/2$& $\pi$  &$\pi$ \\
\hline
\hline
$\Delta_{the}$& $2/3$ &$2/3$ & $2/3$ & $2/3$  & $2/3$& $2/3$ & $2/3$  & $2/3$ & $2/3$ & $2/3$  \\
\hline
$\Delta_{exp}$ &$0.684$ & $0.672$ & $0.689$  &$0.637$   & $0.698$& $0.666$ & $0.703$  &$0.675$& $0.641$  &$0.707$ \\
\hline
\end{tabular}
\end{center}
\label{state}
\end{table*}%

\emph{Experimental prepared states---}In the experiment, we prepared 10 quantum tetrahedron states, which are labeled by ten orange balls on the Bloch sphere in Fig. \ref{exp_state}. Their spherical coordinates and the fluctuation defined in Eq. \eqref{flua} are listed in Table. \ref{state}.

\begin{figure}[!ht]
\includegraphics[width= 8cm]{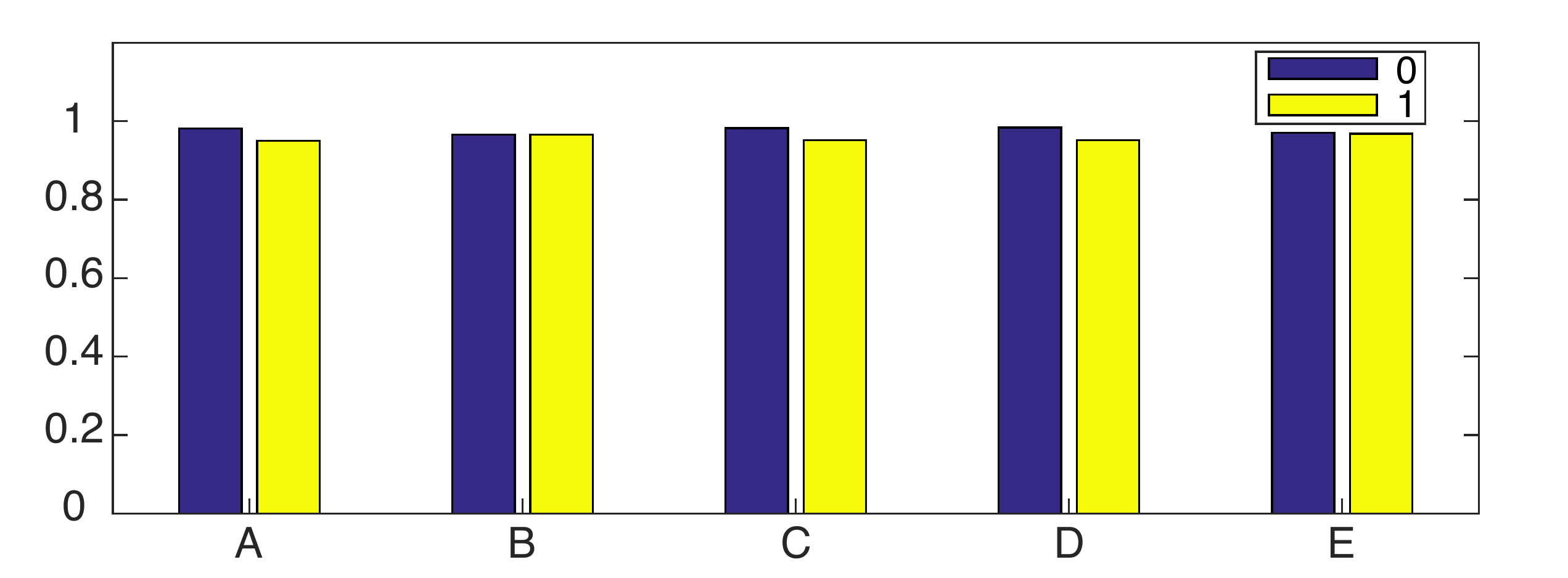}
\caption{Fidelities for the prepared states $\rho^{exp}$ and the theoretical states $\rho^{the}$: A, B,C,D,E combining with the legend 0 and 1, are the same labels as shown in Fig. \ref{exp_state}. These states are different with $\theta$ and $\phi$. }
\label{fidelity}
\end{figure}

To measure the vertex amplitude, we do the full state tomography on our 10 prepared states. We calculated $17$ $2$ms-GRAPE observe  pulses to cover all four-qubit Pauli terms. After that, we calculated the 4-qubit fidelities between all prepared states $\rho^{exp}$ and the theoretical states $\rho^{the}$ with the definition:$F(\rho^{exp},\rho^{the})=trace(\rho^{exp}\rho^{the})/\sqrt{trace(\rho^{exp2}) trace(\rho^{the2})}$. The results are presented as a bar graph shown in Fig.\ref{fidelity}.


\begin{thebibliography}{76}
\expandafter\ifx\csname natexlab\endcsname\relax\def\natexlab#1{#1}\fi
\expandafter\ifx\csname bibnamefont\endcsname\relax
  \def\bibnamefont#1{#1}\fi
\expandafter\ifx\csname bibfnamefont\endcsname\relax
  \def\bibfnamefont#1{#1}\fi
\expandafter\ifx\csname citenamefont\endcsname\relax
  \def\citenamefont#1{#1}\fi
\expandafter\ifx\csname url\endcsname\relax
  \def\url#1{\texttt{#1}}\fi
\expandafter\ifx\csname urlprefix\endcsname\relax\def\urlprefix{URL }\fi
\providecommand{\bibinfo}[2]{#2}
\providecommand{\eprint}[2][]{\url{#2}}

\bibitem[{\citenamefont{Kiefer}(2012)}]{kiefer2012quantum}
\bibinfo{author}{\bibfnamefont{C.}~\bibnamefont{Kiefer}},
  \emph{\bibinfo{title}{Quantum Gravity}}, International Series of Monographs
  on Physics (\bibinfo{publisher}{OUP Oxford}, \bibinfo{year}{2012}), ISBN
  \bibinfo{isbn}{9780191628856}.

\bibitem[{\citenamefont{Rovelli}(2004)}]{book1}
\bibinfo{author}{\bibfnamefont{C.}~\bibnamefont{Rovelli}},
  \emph{\bibinfo{title}{Quantum Gravity}} (\bibinfo{publisher}{Cambridge
  University Press}, \bibinfo{year}{2004}), ISBN \bibinfo{isbn}{1139456156}.

\bibitem[{\citenamefont{Ashtekar}(2005)}]{ashtekar}
\bibinfo{author}{\bibfnamefont{A.}~\bibnamefont{Ashtekar}},
  \bibinfo{journal}{Current Science} \textbf{\bibinfo{volume}{89}},
  \bibinfo{pages}{2064} (\bibinfo{year}{2005}), ISSN \bibinfo{issn}{00113891}.

\bibitem[{\citenamefont{Smolin}(2002)}]{smolin2002three}
\bibinfo{author}{\bibfnamefont{L.}~\bibnamefont{Smolin}},
  \emph{\bibinfo{title}{Three Roads To Quantum Gravity}}, Science masters
  series (\bibinfo{publisher}{Basic Books}, \bibinfo{year}{2002}), ISBN
  \bibinfo{isbn}{9780465078363}.

\bibitem[{\citenamefont{Nicolai}(2014)}]{Nicolai:2013sz}
\bibinfo{author}{\bibfnamefont{H.}~\bibnamefont{Nicolai}},
  \bibinfo{journal}{Fundam. Theor. Phys.} \textbf{\bibinfo{volume}{177}},
  \bibinfo{pages}{369} (\bibinfo{year}{2014}), \eprint{1301.5481}.

\bibitem[{\citenamefont{Polchinski}(1998)}]{polchinski1998string}
\bibinfo{author}{\bibfnamefont{J.}~\bibnamefont{Polchinski}},
  \emph{\bibinfo{title}{String Theory: Volume 1, An Introduction to the Bosonic
  String}}, Cambridge Monographs on Mathematical Physics
  (\bibinfo{publisher}{Cambridge University Press}, \bibinfo{year}{1998}), ISBN
  \bibinfo{isbn}{9781139457408}.

\bibitem[{\citenamefont{Thiemann}(2007)}]{book}
\bibinfo{author}{\bibfnamefont{T.}~\bibnamefont{Thiemann}},
  \emph{\bibinfo{title}{Modern Canonical Quantum General Relativity}}
  (\bibinfo{publisher}{Cambridge University Press}, \bibinfo{year}{2007}).

\bibitem[{\citenamefont{Penrose and Rindler}(1986)}]{penrose1986spinors}
\bibinfo{author}{\bibfnamefont{R.}~\bibnamefont{Penrose}} \bibnamefont{and}
  \bibinfo{author}{\bibfnamefont{W.}~\bibnamefont{Rindler}},
  \emph{\bibinfo{title}{Spinors and Space-Time: Volume 2, Spinor and Twistor
  Methods in Space-Time Geometry}}, Cambridge Monographs on Mathematical
  Physics (\bibinfo{publisher}{Cambridge University Press},
  \bibinfo{year}{1986}), ISBN \bibinfo{isbn}{9780521252676}.

\bibitem[{\citenamefont{Freidel}(2005)}]{Freidel:2005qe}
\bibinfo{author}{\bibfnamefont{L.}~\bibnamefont{Freidel}},
  \bibinfo{journal}{Int. J. Theor. Phys.} \textbf{\bibinfo{volume}{44}},
  \bibinfo{pages}{1769} (\bibinfo{year}{2005}), \eprint{hep-th/0505016}.

\bibitem[{\citenamefont{Loll}(1998)}]{Loll1998}
\bibinfo{author}{\bibfnamefont{R.}~\bibnamefont{Loll}},
  \bibinfo{journal}{Living Reviews in Relativity} \textbf{\bibinfo{volume}{1}},
  \bibinfo{pages}{13} (\bibinfo{year}{1998}), ISSN \bibinfo{issn}{1433-8351}.

\bibitem[{\citenamefont{Niedermaier and Reuter}(2006)}]{Niedermaier2006}
\bibinfo{author}{\bibfnamefont{M.}~\bibnamefont{Niedermaier}} \bibnamefont{and}
  \bibinfo{author}{\bibfnamefont{M.}~\bibnamefont{Reuter}},
  \bibinfo{journal}{Living Reviews in Relativity} \textbf{\bibinfo{volume}{9}},
  \bibinfo{pages}{5} (\bibinfo{year}{2006}), ISSN \bibinfo{issn}{1433-8351}.

\bibitem[{\citenamefont{Penrose}(1971)}]{penroseSN}
\bibinfo{author}{\bibfnamefont{R.}~\bibnamefont{Penrose}},
  \emph{\bibinfo{title}{{Angular momentum: an approach to combinatorial
  spacetime}}} (\bibinfo{publisher}{in T. Bastin (ed.), \emph{Quantum Theory
  and Beyond}, Cambridge University Press}, \bibinfo{year}{1971}).

\bibitem[{\citenamefont{Rovelli and
  Smolin}(1995{\natexlab{a}})}]{Rovelli:1995ac}
\bibinfo{author}{\bibfnamefont{C.}~\bibnamefont{Rovelli}} \bibnamefont{and}
  \bibinfo{author}{\bibfnamefont{L.}~\bibnamefont{Smolin}},
  \bibinfo{journal}{Phys. Rev.} \textbf{\bibinfo{volume}{D52}},
  \bibinfo{pages}{5743} (\bibinfo{year}{1995}{\natexlab{a}}),
  \eprint{gr-qc/9505006}.

\bibitem[{\citenamefont{Rovelli and Smolin}(1988)}]{Rovelli1988}
\bibinfo{author}{\bibfnamefont{C.}~\bibnamefont{Rovelli}} \bibnamefont{and}
  \bibinfo{author}{\bibfnamefont{L.}~\bibnamefont{Smolin}},
  \bibinfo{journal}{Physical Review Letters} \textbf{\bibinfo{volume}{61}},
  \bibinfo{pages}{1155} (\bibinfo{year}{1988}).

\bibitem[{\citenamefont{Ashtekar and Lewandowski}(1993)}]{Ashtekar:1993wf}
\bibinfo{author}{\bibfnamefont{A.}~\bibnamefont{Ashtekar}} \bibnamefont{and}
  \bibinfo{author}{\bibfnamefont{J.}~\bibnamefont{Lewandowski}}
  (\bibinfo{year}{1993}), \eprint{gr-qc/9311010}.

\bibitem[{\citenamefont{Bruegmann et~al.}(1992)\citenamefont{Bruegmann,
  Gambini, and Pullin}}]{Bruegmann:1992gp}
\bibinfo{author}{\bibfnamefont{B.}~\bibnamefont{Bruegmann}},
  \bibinfo{author}{\bibfnamefont{R.}~\bibnamefont{Gambini}}, \bibnamefont{and}
  \bibinfo{author}{\bibfnamefont{J.}~\bibnamefont{Pullin}},
  \bibinfo{journal}{Nucl. Phys.} \textbf{\bibinfo{volume}{B385}},
  \bibinfo{pages}{587} (\bibinfo{year}{1992}), \eprint{hep-th/9202018}.

\bibitem[{\citenamefont{Ashtekar and Lewandowski}(2004)}]{review1}
\bibinfo{author}{\bibfnamefont{A.}~\bibnamefont{Ashtekar}} \bibnamefont{and}
  \bibinfo{author}{\bibfnamefont{J.}~\bibnamefont{Lewandowski}},
  \bibinfo{journal}{Class.Quant.Grav.} \textbf{\bibinfo{volume}{21}},
  \bibinfo{pages}{R53} (\bibinfo{year}{2004}), \eprint{gr-qc/0404018}.

\bibitem[{\citenamefont{Han et~al.}(2007)\citenamefont{Han, Huang, and
  Ma}}]{review}
\bibinfo{author}{\bibfnamefont{M.}~\bibnamefont{Han}},
  \bibinfo{author}{\bibfnamefont{W.}~\bibnamefont{Huang}}, \bibnamefont{and}
  \bibinfo{author}{\bibfnamefont{Y.}~\bibnamefont{Ma}},
  \bibinfo{journal}{Int.J.Mod.Phys.} \textbf{\bibinfo{volume}{D16}},
  \bibinfo{pages}{1397} (\bibinfo{year}{2007}), \eprint{gr-qc/0509064}.

\bibitem[{\citenamefont{Major}(1999)}]{Major:1999md}
\bibinfo{author}{\bibfnamefont{S.~A.} \bibnamefont{Major}},
  \bibinfo{journal}{Am. J. Phys.} \textbf{\bibinfo{volume}{67}},
  \bibinfo{pages}{972} (\bibinfo{year}{1999}), \eprint{gr-qc/9905020}.

\bibitem[{\citenamefont{Han and Hung}(2017)}]{hanhung}
\bibinfo{author}{\bibfnamefont{M.}~\bibnamefont{Han}} \bibnamefont{and}
  \bibinfo{author}{\bibfnamefont{L.-Y.} \bibnamefont{Hung}},
  \bibinfo{journal}{Phys. Rev.} \textbf{\bibinfo{volume}{D95}},
  \bibinfo{pages}{024011} (\bibinfo{year}{2017}), \eprint{1610.02134}.

\bibitem[{\citenamefont{Singh et~al.}(2017)\citenamefont{Singh, McMahon, and
  Brennen}}]{Singh:2017tet}
\bibinfo{author}{\bibfnamefont{S.}~\bibnamefont{Singh}},
  \bibinfo{author}{\bibfnamefont{N.~A.} \bibnamefont{McMahon}},
  \bibnamefont{and} \bibinfo{author}{\bibfnamefont{G.~K.}
  \bibnamefont{Brennen}} (\bibinfo{year}{2017}), \eprint{1702.00392}.

\bibitem[{\citenamefont{Chirco et~al.}(2017)\citenamefont{Chirco, Oriti, and
  Zhang}}]{Chirco:2017vhs}
\bibinfo{author}{\bibfnamefont{G.}~\bibnamefont{Chirco}},
  \bibinfo{author}{\bibfnamefont{D.}~\bibnamefont{Oriti}}, \bibnamefont{and}
  \bibinfo{author}{\bibfnamefont{M.}~\bibnamefont{Zhang}}
  (\bibinfo{year}{2017}), \eprint{1701.01383}.

\bibitem[{\citenamefont{Baez and Muniain}(1994)}]{Baez1994}
\bibinfo{author}{\bibfnamefont{J.~C.} \bibnamefont{Baez}} \bibnamefont{and}
  \bibinfo{author}{\bibfnamefont{J.~P.} \bibnamefont{Muniain}},
  \emph{\bibinfo{title}{{Gauge fields, knots and gravity}}}
  (\bibinfo{publisher}{World Scientific}, \bibinfo{address}{Singapore},
  \bibinfo{year}{1994}).

\bibitem[{\citenamefont{Baez}(1996)}]{Baez:1994hx}
\bibinfo{author}{\bibfnamefont{J.~C.} \bibnamefont{Baez}},
  \bibinfo{journal}{Adv. Math.} \textbf{\bibinfo{volume}{117}},
  \bibinfo{pages}{253} (\bibinfo{year}{1996}), \eprint{gr-qc/9411007}.

\bibitem[{\citenamefont{Oeckl}(2003)}]{Oeckl:2001wm}
\bibinfo{author}{\bibfnamefont{R.}~\bibnamefont{Oeckl}}, \bibinfo{journal}{J.
  Geom. Phys.} \textbf{\bibinfo{volume}{46}}, \bibinfo{pages}{308}
  (\bibinfo{year}{2003}), \eprint{hep-th/0110259}.

\bibitem[{\citenamefont{Gambini and Pullin}(2000)}]{gambini2000loops}
\bibinfo{author}{\bibfnamefont{R.}~\bibnamefont{Gambini}} \bibnamefont{and}
  \bibinfo{author}{\bibfnamefont{J.}~\bibnamefont{Pullin}},
  \emph{\bibinfo{title}{Loops, Knots, Gauge Theories and Quantum Gravity}},
  Cambridge Monographs on Mathematical Physics (\bibinfo{publisher}{Cambridge
  University Press}, \bibinfo{year}{2000}), ISBN \bibinfo{isbn}{9780521654753}.

\bibitem[{\citenamefont{Levin and Wen}(2005)}]{Levin:2004js}
\bibinfo{author}{\bibfnamefont{M.~A.} \bibnamefont{Levin}} \bibnamefont{and}
  \bibinfo{author}{\bibfnamefont{X.-G.} \bibnamefont{Wen}},
  \bibinfo{journal}{Rev. Mod. Phys.} \textbf{\bibinfo{volume}{77}},
  \bibinfo{pages}{871} (\bibinfo{year}{2005}), \eprint{cond-mat/0407140}.

\bibitem[{\citenamefont{Konopka et~al.}(2006)\citenamefont{Konopka,
  Markopoulou, and Smolin}}]{Konopka:2006hu}
\bibinfo{author}{\bibfnamefont{T.}~\bibnamefont{Konopka}},
  \bibinfo{author}{\bibfnamefont{F.}~\bibnamefont{Markopoulou}},
  \bibnamefont{and} \bibinfo{author}{\bibfnamefont{L.}~\bibnamefont{Smolin}}
  (\bibinfo{year}{2006}), \eprint{hep-th/0611197}.

\bibitem[{\citenamefont{{Kirillov}}(2011)}]{2011arXiv1106.6033K}
\bibinfo{author}{\bibfnamefont{A.}~\bibnamefont{{Kirillov}},
  \bibfnamefont{Jr}}, \bibinfo{journal}{ArXiv e-prints}
  (\bibinfo{year}{2011}), \eprint{1106.6033}.

\bibitem[{\citenamefont{Kauffman and Baadhio}(1993)}]{kauffman1993quantum}
\bibinfo{author}{\bibfnamefont{L.}~\bibnamefont{Kauffman}} \bibnamefont{and}
  \bibinfo{author}{\bibfnamefont{R.}~\bibnamefont{Baadhio}},
  \emph{\bibinfo{title}{Quantum Topology}}, Series on Knots and Everything
  (\bibinfo{year}{1993}), ISBN \bibinfo{isbn}{9789814502672}.

\bibitem[{\citenamefont{Turaev and Viro}(1992)}]{Turaev:1992hq}
\bibinfo{author}{\bibfnamefont{V.~G.} \bibnamefont{Turaev}} \bibnamefont{and}
  \bibinfo{author}{\bibfnamefont{O.~Y.} \bibnamefont{Viro}},
  \bibinfo{journal}{Topology} \textbf{\bibinfo{volume}{31}},
  \bibinfo{pages}{865} (\bibinfo{year}{1992}).

\bibitem[{\citenamefont{van~der Veen}(2009)}]{roland1}
\bibinfo{author}{\bibfnamefont{R.}~\bibnamefont{van~der Veen}},
  \bibinfo{journal}{Algebr. Geom. Topol.} \textbf{\bibinfo{volume}{9}},
  \bibinfo{pages}{691} (\bibinfo{year}{2009}), \eprint{0805.0094}.

\bibitem[{\citenamefont{Crane et~al.}(1994)\citenamefont{Crane, Kauffman, and
  Yetter}}]{Crane:1994ji}
\bibinfo{author}{\bibfnamefont{L.}~\bibnamefont{Crane}},
  \bibinfo{author}{\bibfnamefont{L.~H.} \bibnamefont{Kauffman}},
  \bibnamefont{and} \bibinfo{author}{\bibfnamefont{D.~N.} \bibnamefont{Yetter}}
  (\bibinfo{year}{1994}), \eprint{hep-th/9409167}.

\bibitem[{\citenamefont{Rovelli and Smolin}(1995{\natexlab{b}})}]{Rovelli1995}
\bibinfo{author}{\bibfnamefont{C.}~\bibnamefont{Rovelli}} \bibnamefont{and}
  \bibinfo{author}{\bibfnamefont{L.}~\bibnamefont{Smolin}},
  \bibinfo{journal}{Nuclear Physics B} \textbf{\bibinfo{volume}{442}},
  \bibinfo{pages}{593} (\bibinfo{year}{1995}{\natexlab{b}}), ISSN
  \bibinfo{issn}{05503213}.

\bibitem[{\citenamefont{Ashtekar and Lewandowski}(1997)}]{ALarea}
\bibinfo{author}{\bibfnamefont{A.}~\bibnamefont{Ashtekar}} \bibnamefont{and}
  \bibinfo{author}{\bibfnamefont{J.}~\bibnamefont{Lewandowski}},
  \bibinfo{journal}{Class.Quant.Grav.} \textbf{\bibinfo{volume}{14}},
  \bibinfo{pages}{A55} (\bibinfo{year}{1997}), \eprint{gr-qc/9602046}.

\bibitem[{\citenamefont{Ashtekar and Lewandowski}(1998)}]{ALvolume}
\bibinfo{author}{\bibfnamefont{A.}~\bibnamefont{Ashtekar}} \bibnamefont{and}
  \bibinfo{author}{\bibfnamefont{J.}~\bibnamefont{Lewandowski}},
  \bibinfo{journal}{Adv.Theor.Math.Phys.} \textbf{\bibinfo{volume}{1}},
  \bibinfo{pages}{388} (\bibinfo{year}{1998}), \eprint{gr-qc/9711031}.

\bibitem[{\citenamefont{Bianchi}(2009)}]{Bianchi:2008es}
\bibinfo{author}{\bibfnamefont{E.}~\bibnamefont{Bianchi}},
  \bibinfo{journal}{Nucl. Phys.} \textbf{\bibinfo{volume}{B807}},
  \bibinfo{pages}{591} (\bibinfo{year}{2009}), \eprint{0806.4710}.

\bibitem[{\citenamefont{Thiemann}(1998)}]{Thiemann:1996at}
\bibinfo{author}{\bibfnamefont{T.}~\bibnamefont{Thiemann}},
  \bibinfo{journal}{J. Math. Phys.} \textbf{\bibinfo{volume}{39}},
  \bibinfo{pages}{3372} (\bibinfo{year}{1998}), \eprint{gr-qc/9606092}.

\bibitem[{\citenamefont{Ma et~al.}(2010)\citenamefont{Ma, Soo, and
  Yang}}]{Ma:2010fy}
\bibinfo{author}{\bibfnamefont{Y.}~\bibnamefont{Ma}},
  \bibinfo{author}{\bibfnamefont{C.}~\bibnamefont{Soo}}, \bibnamefont{and}
  \bibinfo{author}{\bibfnamefont{J.}~\bibnamefont{Yang}},
  \bibinfo{journal}{Phys. Rev.} \textbf{\bibinfo{volume}{D81}},
  \bibinfo{pages}{124026} (\bibinfo{year}{2010}), \eprint{1004.1063}.

\bibitem[{\citenamefont{Barbieri}(1998)}]{Barbieri:1997ks}
\bibinfo{author}{\bibfnamefont{A.}~\bibnamefont{Barbieri}},
  \bibinfo{journal}{Nucl.Phys.} \textbf{\bibinfo{volume}{B518}},
  \bibinfo{pages}{714} (\bibinfo{year}{1998}), \eprint{gr-qc/9707010}.

\bibitem[{\citenamefont{Baez and Barrett}(1999)}]{Baez:1999tk}
\bibinfo{author}{\bibfnamefont{J.~C.} \bibnamefont{Baez}} \bibnamefont{and}
  \bibinfo{author}{\bibfnamefont{J.~W.} \bibnamefont{Barrett}},
  \bibinfo{journal}{Adv.Theor.Math.Phys.} \textbf{\bibinfo{volume}{3}},
  \bibinfo{pages}{815} (\bibinfo{year}{1999}), \eprint{gr-qc/9903060}.

\bibitem[{\citenamefont{Bianchi et~al.}(2011)\citenamefont{Bianchi, Dona, and
  Speziale}}]{shape}
\bibinfo{author}{\bibfnamefont{E.}~\bibnamefont{Bianchi}},
  \bibinfo{author}{\bibfnamefont{P.}~\bibnamefont{Dona}}, \bibnamefont{and}
  \bibinfo{author}{\bibfnamefont{S.}~\bibnamefont{Speziale}},
  \bibinfo{journal}{Phys.Rev.} \textbf{\bibinfo{volume}{D83}},
  \bibinfo{pages}{044035} (\bibinfo{year}{2011}), \eprint{1009.3402}.

\bibitem[{\citenamefont{Rovelli and Speziale}(2006)}]{Rovelli:2006fw}
\bibinfo{author}{\bibfnamefont{C.}~\bibnamefont{Rovelli}} \bibnamefont{and}
  \bibinfo{author}{\bibfnamefont{S.}~\bibnamefont{Speziale}},
  \bibinfo{journal}{Class. Quant. Grav.} \textbf{\bibinfo{volume}{23}},
  \bibinfo{pages}{5861} (\bibinfo{year}{2006}), \eprint{gr-qc/0606074}.

\bibitem[{\citenamefont{Conrady and Freidel}(2009)}]{CF}
\bibinfo{author}{\bibfnamefont{F.}~\bibnamefont{Conrady}} \bibnamefont{and}
  \bibinfo{author}{\bibfnamefont{L.}~\bibnamefont{Freidel}},
  \bibinfo{journal}{J.Math.Phys.} \textbf{\bibinfo{volume}{50}},
  \bibinfo{pages}{123510} (\bibinfo{year}{2009}), \eprint{0902.0351}.

\bibitem[{\citenamefont{Bianchi and Haggard}(2011)}]{Bianchi:2011ub}
\bibinfo{author}{\bibfnamefont{E.}~\bibnamefont{Bianchi}} \bibnamefont{and}
  \bibinfo{author}{\bibfnamefont{H.~M.} \bibnamefont{Haggard}},
  \bibinfo{journal}{Phys. Rev. Lett.} \textbf{\bibinfo{volume}{107}},
  \bibinfo{pages}{011301} (\bibinfo{year}{2011}), \eprint{1102.5439}.

\bibitem[{\citenamefont{Rovelli and Vidotto}(2014)}]{rovelli2014covariant}
\bibinfo{author}{\bibfnamefont{C.}~\bibnamefont{Rovelli}} \bibnamefont{and}
  \bibinfo{author}{\bibfnamefont{F.}~\bibnamefont{Vidotto}},
  \emph{\bibinfo{title}{Covariant Loop Quantum Gravity: An Elementary
  Introduction to Quantum Gravity and Spinfoam Theory}}, Cambridge Monographs
  on Mathematical Physics (\bibinfo{publisher}{Cambridge University Press},
  \bibinfo{year}{2014}), ISBN \bibinfo{isbn}{9781107069626}.

\bibitem[{\citenamefont{Perez}(2013)}]{Perez2012}
\bibinfo{author}{\bibfnamefont{A.}~\bibnamefont{Perez}},
  \bibinfo{journal}{Living Rev.Rel.} \textbf{\bibinfo{volume}{16}},
  \bibinfo{pages}{3} (\bibinfo{year}{2013}), \eprint{1205.2019}.

\bibitem[{\citenamefont{Ooguri}(1992)}]{Ooguri:1992eb}
\bibinfo{author}{\bibfnamefont{H.}~\bibnamefont{Ooguri}},
  \bibinfo{journal}{Mod. Phys. Lett.} \textbf{\bibinfo{volume}{A7}},
  \bibinfo{pages}{2799} (\bibinfo{year}{1992}), \eprint{hep-th/9205090}.

\bibitem[{\citenamefont{Barrett and Crane}(1998)}]{BC}
\bibinfo{author}{\bibfnamefont{J.~W.} \bibnamefont{Barrett}} \bibnamefont{and}
  \bibinfo{author}{\bibfnamefont{L.}~\bibnamefont{Crane}},
  \bibinfo{journal}{J.Math.Phys.} \textbf{\bibinfo{volume}{39}},
  \bibinfo{pages}{3296} (\bibinfo{year}{1998}), \eprint{gr-qc/9709028}.

\bibitem[{\citenamefont{Engle et~al.}(2008)\citenamefont{Engle, Livine,
  Pereira, and Rovelli}}]{EPRL}
\bibinfo{author}{\bibfnamefont{J.}~\bibnamefont{Engle}},
  \bibinfo{author}{\bibfnamefont{E.}~\bibnamefont{Livine}},
  \bibinfo{author}{\bibfnamefont{R.}~\bibnamefont{Pereira}}, \bibnamefont{and}
  \bibinfo{author}{\bibfnamefont{C.}~\bibnamefont{Rovelli}},
  \bibinfo{journal}{Nucl.Phys.} \textbf{\bibinfo{volume}{B799}},
  \bibinfo{pages}{136} (\bibinfo{year}{2008}), \eprint{0711.0146}.

\bibitem[{\citenamefont{Freidel and Krasnov}(2008)}]{FK}
\bibinfo{author}{\bibfnamefont{L.}~\bibnamefont{Freidel}} \bibnamefont{and}
  \bibinfo{author}{\bibfnamefont{K.}~\bibnamefont{Krasnov}},
  \bibinfo{journal}{Class.Quant.Grav.} \textbf{\bibinfo{volume}{25}},
  \bibinfo{pages}{125018} (\bibinfo{year}{2008}), \eprint{0708.1595}.

\bibitem[{\citenamefont{Kaminski et~al.}(2010)\citenamefont{Kaminski,
  Kisielowski, and Lewandowski}}]{KKL}
\bibinfo{author}{\bibfnamefont{W.}~\bibnamefont{Kaminski}},
  \bibinfo{author}{\bibfnamefont{M.}~\bibnamefont{Kisielowski}},
  \bibnamefont{and}
  \bibinfo{author}{\bibfnamefont{J.}~\bibnamefont{Lewandowski}},
  \bibinfo{journal}{Class. Quant. Grav.} \textbf{\bibinfo{volume}{27}},
  \bibinfo{pages}{095006} (\bibinfo{year}{2010}), \eprint{0909.0939}.

\bibitem[{\citenamefont{Noui and Roche}(2003)}]{NP}
\bibinfo{author}{\bibfnamefont{K.}~\bibnamefont{Noui}} \bibnamefont{and}
  \bibinfo{author}{\bibfnamefont{P.}~\bibnamefont{Roche}},
  \bibinfo{journal}{Class.Quant.Grav.} \textbf{\bibinfo{volume}{20}},
  \bibinfo{pages}{3175} (\bibinfo{year}{2003}), \eprint{gr-qc/0211109}.

\bibitem[{\citenamefont{Han and Thiemann}(2013)}]{HT}
\bibinfo{author}{\bibfnamefont{M.}~\bibnamefont{Han}} \bibnamefont{and}
  \bibinfo{author}{\bibfnamefont{T.}~\bibnamefont{Thiemann}},
  \bibinfo{journal}{Class. Quant. Grav.} \textbf{\bibinfo{volume}{30}},
  \bibinfo{pages}{235024} (\bibinfo{year}{2013}), \eprint{1010.5444}.

\bibitem[{\citenamefont{Livine and Speziale}(2007)}]{LS}
\bibinfo{author}{\bibfnamefont{E.~R.} \bibnamefont{Livine}} \bibnamefont{and}
  \bibinfo{author}{\bibfnamefont{S.}~\bibnamefont{Speziale}},
  \bibinfo{journal}{Phys.Rev.} \textbf{\bibinfo{volume}{D76}},
  \bibinfo{pages}{084028} (\bibinfo{year}{2007}), \eprint{0705.0674}.

\bibitem[{\citenamefont{Dupuis et~al.}(2012)\citenamefont{Dupuis, Freidel,
  Livine, and Speziale}}]{DFLS}
\bibinfo{author}{\bibfnamefont{M.}~\bibnamefont{Dupuis}},
  \bibinfo{author}{\bibfnamefont{L.}~\bibnamefont{Freidel}},
  \bibinfo{author}{\bibfnamefont{E.~R.} \bibnamefont{Livine}},
  \bibnamefont{and} \bibinfo{author}{\bibfnamefont{S.}~\bibnamefont{Speziale}},
  \bibinfo{journal}{J.Math.Phys.} \textbf{\bibinfo{volume}{53}},
  \bibinfo{pages}{032502} (\bibinfo{year}{2012}), \eprint{1107.5274}.

\bibitem[{\citenamefont{Ding et~al.}(2011)\citenamefont{Ding, Han, and
  Rovelli}}]{generalize}
\bibinfo{author}{\bibfnamefont{Y.}~\bibnamefont{Ding}},
  \bibinfo{author}{\bibfnamefont{M.}~\bibnamefont{Han}}, \bibnamefont{and}
  \bibinfo{author}{\bibfnamefont{C.}~\bibnamefont{Rovelli}},
  \bibinfo{journal}{Phys.Rev.} \textbf{\bibinfo{volume}{D83}},
  \bibinfo{pages}{124020} (\bibinfo{year}{2011}), \eprint{1011.2149}.

\bibitem[{\citenamefont{Haggard et~al.}(2015)\citenamefont{Haggard, Han,
  Kaminski, and Riello}}]{HHKR}
\bibinfo{author}{\bibfnamefont{H.~M.} \bibnamefont{Haggard}},
  \bibinfo{author}{\bibfnamefont{M.}~\bibnamefont{Han}},
  \bibinfo{author}{\bibfnamefont{W.}~\bibnamefont{Kaminski}}, \bibnamefont{and}
  \bibinfo{author}{\bibfnamefont{A.}~\bibnamefont{Riello}},
  \bibinfo{journal}{Nucl. Phys.} \textbf{\bibinfo{volume}{B900}},
  \bibinfo{pages}{1} (\bibinfo{year}{2015}), \eprint{1412.7546}.

\bibitem[{\citenamefont{Rovelli}(2006)}]{propagator}
\bibinfo{author}{\bibfnamefont{C.}~\bibnamefont{Rovelli}},
  \bibinfo{journal}{Phys.Rev.Lett.} \textbf{\bibinfo{volume}{97}},
  \bibinfo{pages}{151301} (\bibinfo{year}{2006}), \eprint{gr-qc/0508124}.

\bibitem[{\citenamefont{Barrett
  et~al.}(2010{\natexlab{a}})\citenamefont{Barrett, Dowdall, Fairbairn,
  Hellmann, and Pereira}}]{semiclassical}
\bibinfo{author}{\bibfnamefont{J.~W.} \bibnamefont{Barrett}},
  \bibinfo{author}{\bibfnamefont{R.}~\bibnamefont{Dowdall}},
  \bibinfo{author}{\bibfnamefont{W.~J.} \bibnamefont{Fairbairn}},
  \bibinfo{author}{\bibfnamefont{F.}~\bibnamefont{Hellmann}}, \bibnamefont{and}
  \bibinfo{author}{\bibfnamefont{R.}~\bibnamefont{Pereira}},
  \bibinfo{journal}{Class.Quant.Grav.} \textbf{\bibinfo{volume}{27}},
  \bibinfo{pages}{165009} (\bibinfo{year}{2010}{\natexlab{a}}),
  \eprint{0907.2440}.

\bibitem[{\citenamefont{Barrett
  et~al.}(2010{\natexlab{b}})\citenamefont{Barrett, Fairbairn, and
  Hellmann}}]{Barrett:2009as}
\bibinfo{author}{\bibfnamefont{J.~W.} \bibnamefont{Barrett}},
  \bibinfo{author}{\bibfnamefont{W.~J.} \bibnamefont{Fairbairn}},
  \bibnamefont{and} \bibinfo{author}{\bibfnamefont{F.}~\bibnamefont{Hellmann}},
  \bibinfo{journal}{Int. J. Mod. Phys.} \textbf{\bibinfo{volume}{A25}},
  \bibinfo{pages}{2897} (\bibinfo{year}{2010}{\natexlab{b}}),
  \eprint{0912.4907}.

\bibitem[{\citenamefont{Conrady and Freidel}(2008)}]{CFsemiclassical}
\bibinfo{author}{\bibfnamefont{F.}~\bibnamefont{Conrady}} \bibnamefont{and}
  \bibinfo{author}{\bibfnamefont{L.}~\bibnamefont{Freidel}},
  \bibinfo{journal}{Phys.Rev.} \textbf{\bibinfo{volume}{D78}},
  \bibinfo{pages}{104023} (\bibinfo{year}{2008}), \eprint{0809.2280}.

\bibitem[{\citenamefont{Han and Zhang}(2013)}]{HZ}
\bibinfo{author}{\bibfnamefont{M.}~\bibnamefont{Han}} \bibnamefont{and}
  \bibinfo{author}{\bibfnamefont{M.}~\bibnamefont{Zhang}},
  \bibinfo{journal}{Class.Quant.Grav.} \textbf{\bibinfo{volume}{30}},
  \bibinfo{pages}{165012} (\bibinfo{year}{2013}), \eprint{1109.0499}.

\bibitem[{\citenamefont{Bianchi et~al.}(2009)\citenamefont{Bianchi, Magliaro,
  and Perini}}]{propagator2}
\bibinfo{author}{\bibfnamefont{E.}~\bibnamefont{Bianchi}},
  \bibinfo{author}{\bibfnamefont{E.}~\bibnamefont{Magliaro}}, \bibnamefont{and}
  \bibinfo{author}{\bibfnamefont{C.}~\bibnamefont{Perini}},
  \bibinfo{journal}{Nucl.Phys.} \textbf{\bibinfo{volume}{B822}},
  \bibinfo{pages}{245} (\bibinfo{year}{2009}), \eprint{0905.4082}.

\bibitem[{\citenamefont{Bianchi and Ding}(2012)}]{propagator3}
\bibinfo{author}{\bibfnamefont{E.}~\bibnamefont{Bianchi}} \bibnamefont{and}
  \bibinfo{author}{\bibfnamefont{Y.}~\bibnamefont{Ding}},
  \bibinfo{journal}{Phys.Rev.} \textbf{\bibinfo{volume}{D86}},
  \bibinfo{pages}{104040} (\bibinfo{year}{2012}), \eprint{1109.6538}.

\bibitem[{\citenamefont{Magliaro and Perini}(2011)}]{claudio1}
\bibinfo{author}{\bibfnamefont{E.}~\bibnamefont{Magliaro}} \bibnamefont{and}
  \bibinfo{author}{\bibfnamefont{C.}~\bibnamefont{Perini}},
  \bibinfo{journal}{Europhys.Lett.} \textbf{\bibinfo{volume}{95}},
  \bibinfo{pages}{30007} (\bibinfo{year}{2011}), \eprint{1108.2258}.

\bibitem[{\citenamefont{Alesci}(2009)}]{Alesci:2009ys}
\bibinfo{author}{\bibfnamefont{E.}~\bibnamefont{Alesci}}, in
  \emph{\bibinfo{booktitle}{{3rd Stueckelberg Workshop on Relativistic Field
  Theories Pescara, Italy, July 8-18, 2008}}} (\bibinfo{year}{2009}),
  \eprint{0903.4329}.

\bibitem[{\citenamefont{Rovelli}(2011)}]{Rovelli:2010vv}
\bibinfo{author}{\bibfnamefont{C.}~\bibnamefont{Rovelli}}, \bibinfo{journal}{J.
  Phys. Conf. Ser.} \textbf{\bibinfo{volume}{314}}, \bibinfo{pages}{012006}
  (\bibinfo{year}{2011}), \eprint{1010.1939}.

\bibitem[{\citenamefont{Han}(2014)}]{lowE}
\bibinfo{author}{\bibfnamefont{M.}~\bibnamefont{Han}},
  \bibinfo{journal}{Phys.Rev.} \textbf{\bibinfo{volume}{D89}},
  \bibinfo{pages}{124001} (\bibinfo{year}{2014}), \eprint{1308.4063}.

\bibitem[{\citenamefont{Han}(2017)}]{Han:2017xwo}
\bibinfo{author}{\bibfnamefont{M.}~\bibnamefont{Han}} (\bibinfo{year}{2017}),
  \eprint{1705.09030}.

\bibitem[{\citenamefont{Minkowski}(1989)}]{Minkowski}
\bibinfo{author}{\bibfnamefont{H.}~\bibnamefont{Minkowski}},
  \emph{\bibinfo{title}{{Ausgew\"{a}hlte Arbeiten zur Zahlentheorie und zur
  Geometrie}}}, vol.~\bibinfo{volume}{12} of
  \emph{\bibinfo{series}{Teubner-Archiv zur Mathematik}}
  (\bibinfo{publisher}{Springer Vienna}, \bibinfo{address}{Vienna},
  \bibinfo{year}{1989}), ISBN \bibinfo{isbn}{978-3-211-95845-2},
  \urlprefix\url{http://www.springerlink.com/index/10.1007/978-3-7091-9536-9}.

\bibitem[{\citenamefont{Han and Zhang}(2016)}]{Han:2016fgh}
\bibinfo{author}{\bibfnamefont{M.}~\bibnamefont{Han}} \bibnamefont{and}
  \bibinfo{author}{\bibfnamefont{M.}~\bibnamefont{Zhang}},
  \bibinfo{journal}{Phys. Rev.} \textbf{\bibinfo{volume}{D94}},
  \bibinfo{pages}{104075} (\bibinfo{year}{2016}), \eprint{1606.02826}.

\bibitem[{\citenamefont{Bianchi et~al.}(2006)\citenamefont{Bianchi, Modesto,
  Rovelli, and Speziale}}]{propagator1}
\bibinfo{author}{\bibfnamefont{E.}~\bibnamefont{Bianchi}},
  \bibinfo{author}{\bibfnamefont{L.}~\bibnamefont{Modesto}},
  \bibinfo{author}{\bibfnamefont{C.}~\bibnamefont{Rovelli}}, \bibnamefont{and}
  \bibinfo{author}{\bibfnamefont{S.}~\bibnamefont{Speziale}},
  \bibinfo{journal}{Class.Quant.Grav.} \textbf{\bibinfo{volume}{23}},
  \bibinfo{pages}{6989} (\bibinfo{year}{2006}), \eprint{gr-qc/0604044}.

\bibitem[{\citenamefont{Ashtekar}(1986)}]{Ashtekar:1986yd}
\bibinfo{author}{\bibfnamefont{A.}~\bibnamefont{Ashtekar}},
  \bibinfo{journal}{Phys. Rev. Lett.} \textbf{\bibinfo{volume}{57}},
  \bibinfo{pages}{2244} (\bibinfo{year}{1986}).

\bibitem[{\citenamefont{Barbero~G.}(1995)}]{barbero}
\bibinfo{author}{\bibfnamefont{J.~F.} \bibnamefont{Barbero~G.}},
  \bibinfo{journal}{Phys.Rev.} \textbf{\bibinfo{volume}{D51}},
  \bibinfo{pages}{5507} (\bibinfo{year}{1995}), \eprint{gr-qc/9410014}.

\bibitem[{\citenamefont{Ashtekar et~al.}(1998)\citenamefont{Ashtekar, Corichi,
  and Zapata}}]{Ashtekar:1998ak}
\bibinfo{author}{\bibfnamefont{A.}~\bibnamefont{Ashtekar}},
  \bibinfo{author}{\bibfnamefont{A.}~\bibnamefont{Corichi}}, \bibnamefont{and}
  \bibinfo{author}{\bibfnamefont{J.~A.} \bibnamefont{Zapata}},
  \bibinfo{journal}{Class. Quant. Grav.} \textbf{\bibinfo{volume}{15}},
  \bibinfo{pages}{2955} (\bibinfo{year}{1998}), \eprint{gr-qc/9806041}.

\end{thebibliography}
\end{document}